\documentclass[a4paper,11pt]{article}

\usepackage{lineno}

\usepackage[utf8]{inputenc}
\usepackage{amsmath,amsthm,amsfonts,amssymb,amscd}
\usepackage{bbm}
\usepackage{multirow,booktabs}
\usepackage{lastpage}
\usepackage{enumitem}
\usepackage{fancyhdr}
\usepackage{mathrsfs}
\usepackage{wrapfig}
\usepackage{setspace}
\usepackage{calc}
\usepackage{multicol}
\usepackage{cancel}
\usepackage{graphicx,bm,appendix}
\usepackage[most]{tcolorbox}
\usepackage{ upgreek }

\usepackage{jheppub} 

\newcommand{\nocontentsline}[3]{}
\let\origcontentsline\addcontentsline
\newcommand\stoptoc{\let\addcontentsline\nocontentsline}
\newcommand\resumetoc{\let\addcontentsline\origcontentsline}

\newcommand{\ADSxS}{AdS$_3\times$S$^1$\ }

\newcommand{\mD}{\mathcal{D}}

\newcommand{\mK}{\mathcal{K}}
\newcommand{\mL}{\mathcal{L}}
\newcommand{\mO}{\mathcal{O}}
\newcommand{\mV}{\mathcal{V}}
\newcommand{\nablaII}{\bar{\nabla}}
\newcommand{\boxII}{\bar{\square}}

\newcommand{\Poincare}{Poincar\'e\ }
\newcommand{\Tr}{\text{tr}}

\newtheorem{defn}{Box}

\newcommand{\BigBox}[2]{
\begin{tcolorbox}[colback=black!5!white,colframe=white!50!black,title={#1}]
#2
\end{tcolorbox}
}

\newcommand{\ADS}[1]{
\begin{tcolorbox}[colback=green!5!white,colframe=green!50!black,title=AdS$_3\times $S$^1$ example:]
#1
\end{tcolorbox}
}

\title{\boldmath Scale without Conformal Invariance in bottom-up Holography}

\author{Lavish Chawla,}
\author{Mario Flory}
\affiliation{Institute of Theoretical Physics, Jagiellonian University, \\
{\L}ojasiewicza 11, 
30-348 Krak{\'o}w, Poland}

\emailAdd{lavish.lavish@doctoral.uj.edu.pl}
\emailAdd{mflory@th.if.uj.edu.pl}

\abstract{
In holography, the isometry group of the bulk spacetime corresponds to the symmetries of the boundary theory. We thus approach the question of whether (and when) scale invariance in combination with \Poincare invariance implies full conformal invariance in quantum field theory from a holographic bulk perspective. To do so, we study bulk spacetimes that include a warped extra dimension and in which the isometry group corresponds to scale without conformal invariance. Firstly, we show that the bulk Weyl tensor plays a pivotal role in distinguishing those metrics exhibiting conformal invariance (Weyl=0) from those merely exhibiting scale invariance (Weyl$\neq$0). Based on this, we then prove the following theorem: For putative boundary theories with $n\geq2$ dimensions, the bulk metric can not exhibit scale without conformal invariance if its warped extra dimension is compact and the null energy condition is required to hold. For $n=1$, we discuss that a more general ansatz for the bulk metric must be made, a detailed analysis of which is left for future research. 
\\

\noindent
A 12-minute video abstract can be found at \href{https://youtu.be/J_jHazxEvVk}{this link}.
}

\begin{document}
\maketitle
\flushbottom

\section{Introduction and Summary of Results}
\label{sec:intro}

Over the course of the last few decades, the \textit{Anti-de Sitter/Conformal Field Theory (AdS/CFT) correspondence} \cite{Maldacena:1997re} has proven to be a powerful and versatile instrument in the study of strongly coupled quantum field theories, which has led to many fascinating insights (see \cite{Ammon:2015wua} for an overview). At the same time, this progress has also increased interest and opened new avenues in the study of CFTs from a purely field theory perspective. However, one old and yet partially unsolved problem persists at the very heart of the study of CFTs: The question of when the terms \textit{scale invariance} and \textit{conformal invariance} deserve to be treated as (almost-) synonyms in relativistic field theories (see \cite{Nakayama:2013is} for a review).  Simply speaking, a CFT is a theory that, in addition to the \Poincare invariance of relativistic quantum field theory, is also invariant under scale transformations and special conformal transformations. The conformal algebra explicitly reads:
\begin{align}
\tcbhighmath[colback=green!0!white,colframe=green!75!black,size=small]{
  \begin{aligned}
    \text{Conformal invariance:}\hspace{5.95cm}
  \\
\tcbhighmath[colback=teal!0!white,colframe=teal!100!white,size=small]{
  \begin{aligned}
  \text{Scale without conformal invariance (SwCI):}\hspace{2.05cm}
  \\
\tcbhighmath[colback=blue!0!white,colframe=blue!100!white,size=small]{
\begin{aligned}
&\text{Poincar\'e invariance:}
\\
& \relax   [M_{ab}, M_{cd}] = \eta_{ac} M_{db} - \eta_{bc} M_{da} - \eta_{ad} M_{cb} + \eta_{bd} M_{ca}  \\
 & \relax \ \   [P_a, M_{bd}] = 2\eta_{a[b}P_{d]}
  \end{aligned}
}\\
  \begin{aligned}
\relax [D, P_a] &= P_a   \hspace{5.55cm}
  \end{aligned}
    \end{aligned}
}\\
\relax [K_a, M_{bd}] = 2 \, \eta_{a[b}K_{d]} \hspace{5.12cm}
\\
\relax [P_a, K_b] = 2\, (\eta_{ab} D + M_{ab}) \hspace{4.02cm}
\\
\relax [D, K_a] = -K_a \hspace{5.75cm}
    \end{aligned}
    }
    \label{algebras}
\end{align}
and all other brackets not explicitly mentioned vanish. Here, the $M_{ab}$ generate boosts, the $P_a$ generate translations, the operator $D$ generates scale transformations (also called dilatations), and the $K_a$ are the generators of special conformal transformations.\footnote{For an $n$-dimensional field theory in Minkowski space $\eta_{ab}$, we assume indices $a,b,...$ to run from $0$ to $n-1$. We are following here the conventions of \cite{Freedman:2012zz}, as we will express these generators and later as Killing vectors of a curved spacetime, which we would like to be manifestly real. }
The problem alluded to above now becomes very clear: Conformal invariance is an extension of \Poincare invariance, but as indicated in \eqref{algebras} there is a closed subalgebra that lies in between them. \textit{Scale without conformal invariance (SwCI)} is a symmetry that includes invariance under scale transformations, but not invariance under special conformal transformations. One would hence expect that it should be easy to construct healthy quantum field theories that exhibit SwCI instead of full conformal invariance, but this is indeed a very non-trivial task. 
As we will explain in our literature overview section \ref{sec::Overview}, the possibility of physical theories exhibiting SwCI depends strongly on the dimension $n$ of the field theory. While there exists a clear and famous no-go theorem forbidding SwCI in $n=2$ dimensions \cite{Polchinski:1987dy} and a host of arguments for the same conclusion in $n=4$ \cite{MACK1969174,CALLAN197042,PhysRevD.2.753,Zheng:2011bp,Antoniadis:2011gn,Nakayama:2011tk,Nakayama:2011wq,Fortin:2011sz,Fortin:2012hn,Luty:2012ww,Nakayama:2012nd,
Dymarsky:2013pqa,Yonekura:2014tha,Nakayama:2020bsx}, theories exhibiting SwCI may exist in $n=3$ or $n\geq5$, but appear to be rare or exceptional \cite{Jackiw:2011vz,El-Showk:2011xbs}. The complicated history of this research topic is reviewed in \cite{Nakayama:2013is}.
\\

Our goal will be to approach the issue of scale vs.~conformal invariance from the point of view of gauge/gravity duality, and bottom-up holographic model building. The philosophy behind this is simple: In gauge/gravity duality, important insights on the field theory side tend to correspond to important (but entirely different) insights on the gravity side. A good example for that is probably the famous realisation that the first law of entanglement in quantum field theory corresponds to Einstein's equations for the dynamical bulk metric, assuming the validity of the Ryu-Takayanagi formula \cite{Lashkari:2013koa,Faulkner:2013ica}. 
If there is some law of nature or mathematical physics that prevents field theories living on flat space from exhibiting SwCI, we should then naively expect that there is some theorem that can be proven by the methods of gravitational physics and differential geometry forbidding the existence of geometries that would be holographically dual to such field theories. However, this naive expectation is not guaranteed, and all logical possibilities are summarised in table \ref{tab::T1}. We claim that all possibilities listed in table \ref{tab::T1} would be interesting in their own way, and so the holographic approach to the problem of scale vs.~conformal invariance is a worthwhile undertaking. 
\\

\begin{table}
    \centering
    \begin{tabular}{c|c|c|c}

Case & Does QFT   &  Does the bulk allow  & Would this be \\
 &  allow SwCI?   &  allow SwCI?  & interesting? \\[6pt]
\hline
\hline
 1 &  Yes & Yes & Yes, holographic model building! \\[6pt] \hline
 2 &  No  & No  & Yes, holographic no-go theorems! \\[6pt] \hline
 3 &  Yes & No  & Yes, holographic theories are special! \\[6pt] \hline
 4 &  No  & Yes & Yes, exploration of forbidden  \\
   &      &  & landscape of holography 
    \end{tabular}
    \caption{As we will see in section \ref{sec::Overview}, whether quantum field theories (QFTs) are allowed to exhibit SwCI depends strongly on their dimension $n$, and the exact requirements that are imposed on a theory to be physical. Whether (and under what conditions) holographic bottom-up models of SwCI exist is the subject of this study. There are four possible cases: If both sides of a putative holographic duality allow for SwCI (case 1), then we can use holography to try and construct explicit examples of such a rare type of behaviour. If neither side allows SwCI (case 2), then there has to be some kind of gravitational no-go theorem that can be proven. If the two sides disagree, then either theories with a holographic dual are special (compared to generic QFTs), or we have discovered a "forbidden landscape" within holography \cite{Nakayama:2009qu,Nakayama:2009fe} (cases 3 and 4 respectively).
}
   \label{tab::T1}
\end{table}

Following in the footsteps of earlier papers \cite{Nakayama:2009fe,Nakayama:2010zz,Balasubramanian:2013ux,Flory:2017mal,Rykala} (in section \ref{sec::Overview} we give a more detailed discussion of the literature) we will thus present an ansatz in section \ref{sec::WPA} for a bulk metric of the form
\begin{align}
ds^2=g_{\mu\nu}dx^\mu dx^\nu = S(w,\theta) \underbrace{\eta_{ab} dx^a dx^b}_{\text{fiber}} + \underbrace{h_{ij}dy^i dy^j}_{\text{base space}}
\label{Metric_ansatz}
\end{align}
whose holographically dual field theory could exhibit SwCI (if it exists at all). This ansatz \eqref{Metric_ansatz} includes $d$ bulk dimensions, specifically $n=d-2$ corresponding to a Minkowski fiber (ensuring \Poincare invariance), one extended extra dimension (corresponding to the radial direction in AdS/CFT), and one warped compact extra dimension. Throughout section \ref{sec::WPA}, we will discuss this ansatz in detail, including its basic geometry (section \ref{sec::AnsatzAndNotation}), its isometry group and Killing equations (section \ref{sec::KillingAlgebra}), its topological properties as a consequence of the compact extra dimension (section \ref{sec::Topology}), how these topics intersect in a non-trivial manner (section \ref{sec::GBandCurvature}), and how the \textit{null energy condition (NEC)} can be formulated in such a background (section \ref{sec::NEC}). 
\\

Our paper achieves two main results. The first one concerns the distinction between scale and conformal invariance in models of the type we are investigating. These models \textit{may} exhibit SwCI, but they also contain a subset where scale invariance is extended to full conformal invariance -- AdS$_{d-1}\times$S${}^1$ is the simplest example. As we show in section \ref{sec::Weyl}, the bulk Weyl tensor $C_{\alpha\beta\gamma\delta}$ serves to distinguish these two cases:
\BigBox{
\begin{defn}\label{B0} Role of the Weyl-tensor $C$\end{defn}
}{
$    C_{\alpha\beta\gamma\delta}=0 \Leftrightarrow \text{scale invariance is extended to full conformal invariance.}$
}
\noindent
Some basic results about the form of the Weyl tensor and the conditions for the extension to full conformal invariance are given in sections \ref{sec::Weyl_form} and \ref{sec::ConformalConditions} respectively. In section \ref{sec::CIphi0}, we prove the $\Leftarrow$ direction of the above statement\footnote{The $\Leftarrow$ direction of this statement was already shown in \cite{Rykala} using a more complicated method.}, and in section \ref{sec::phi0impliesconformal} the $\Rightarrow$ direction. Section \ref{sec::Weyl} ends with a discussion of how a vanishing of the Weyl tensor relates to the NEC (section \ref{sec::det_phi}), and the explicit Petrov classification of the metric ansatz (section \ref{sec::Petrov}). 
Our second main result is the proof of the following conjecture in section \ref{sec::proofs}:
\BigBox{
\begin{defn}\label{B1} Main Conjecture\end{defn}
}{
For a metric of the form \eqref{Metric_ansatz} to be a physical bulk-model of SwCI, we demand the following three conditions to hold:
\begin{itemize}
\item \textbf{Local geometrical condition:} The Killing equations allow for Killing vectors forming the SwCI algebra in \eqref{algebras} to exist, but not any additional Killing vectors that would extend SwCI to full conformal invariance. 
\item \textbf{Topological condition:} The warped extra dimension is compact, i.e.~the base space with metric $h_{ij}$ has the topology of a cylinder. 
\item \textbf{Physical condition:} The null energy condition (NEC) holds everywhere in the spacetime.
\end{itemize}
Our main conjecture is that \textbf{these three conditions are mutually exclusive}.
}
\noindent
This conjecture is motivated by earlier work \cite{Flory:2017mal,Rykala}. Section \ref{sec::beta_scaling} introduces and investigates a certain scaling-freedom that is very useful in understanding the problem at hand. Section \ref{sec::special_cases} provides several short and elegant proofs of the main conjecture in box \ref{B1} which are valid only under special simplifying conditions. A generic, but also more technical proof is then finally provided in section \ref{sec::full_proof}. 
\\

By proving the conjecture of box \ref{B1}, we thus establish a relatively general holographic no-go theorem against SwCI, going beyond what existed in the earlier literature \cite{Nakayama:2009fe,Nakayama:2010zz,Balasubramanian:2013ux,Flory:2017mal,Rykala}: \textit{No Field Theory that has a holographically dual description in terms of classical Einstein gravity with matter that satisfies the NEC and one compact extra dimension can exhibit SwCI.} In the language of table \ref{tab::T1}, our bulk results hence correspond to cases 2 or 3, depending on the status of SwCI in the field theory side (and thus they are interesting). More concretely, as said above and in section \ref{sec::Overview}, no-go theorems on the field theory side exist in $n=2$ and $n=4$, while examples of physical theories exhibiting SwCI exist in $n=3$ and $n\geq 5$. It hence appears that in the latter case, theories with a simple holographic description are special with respect to SwCI (case 3). Notably, the examples where SwCI is realised that we are about to discuss in the next section are both \textit{free} theories, and we are not aware of any such examples that are strongly coupled. As strong coupling and large $N$ limits are typical for examples of theories with a classical Einsteinian holographic dual, one may wonder whether a field theory no-go theorem exists for general $n$ once the assumption of strong interactions is added as a condition. See also the comments in footnote \ref{fn5} below.  
\\

The main text closes with a discussion of several avenues in which our results may be generalised in the future. This includes the possibility of changing the topology of the warped extra dimension (section \ref{sec::outlook_topology}), the relation between our type of overall approach and the so-called "Synge g-method" (section \ref{sec::Synge}), the possibility of relaxing the NEC (section \ref{sec::outlook_ECs}), and the issues surrounding the number of bulk dimensions (section \ref{sec::dimensionality}). 
Some technical results are given in the appendices, the arguably most important of which are appendices \ref{sec::1Dfiber} and \ref{sec::higherDfiber}. Appendix \ref{sec::higherDfiber} makes it explicit that our results apply for general dimension $d\geq4$, after for most of the paper we have limited ourselves to the $d=4$ case for the sake of simplicity. Appendix \ref{sec::1Dfiber} deals with the case $d=3$, which would correspond to a $n=1$ dimensional dual field theory. As we explain, a more general ansatz for the bulk metric would be needed in this case, which we however do not investigate further in this manuscript. This might be an interesting topic to study in the future.

\section{Motivation and Literature Overview}
\label{sec::Overview}

In the literature on relativistic quantum field theory, the distinction between conformal invariance and scale \textit{without} conformal invariance (SwCI), evident from the algebra \eqref{algebras}\footnote{Throughout this paper, when talking about scale and conformal invariance, we always consider them to be in combination with \Poincare invariance as in \eqref{algebras}. See \cite{Nakayama:2013is,Farnsworth:2024iwc,Awad:2000ie,Awad:2000ac,Nakayama:2009ww,Nakayama:2012ed} for comments and some literature references concerning scale and conformal invariance in combination with other spacetime symmetries.  See \cite{PhysRevD.5.2519,PhysRevD.10.480} for some subtleties concerning the interplay between large (i.e.~non-infinitesimal) conformal transformations and the causal ordering of points in Minkowski space.}, has been known since a long while, see e.g.~\cite{Coleman:1970je} for a nice historical overview. 
However, as in practice scale invariance is often extended to full conformal invariance, this subtle distinction has sometimes been forgotten, as illustrated by the famous anecdote about an argument taking place at a physics conference in Dubna, retold at the beginning of the beautiful review paper \cite{Nakayama:2013is}. 
But what is the cause for this extension to take place so reliably? 
 
Concrete mathematical arguments are surprisingly hard to formulate. The most famous result exists for $n=2$ dimensional field theories \cite{Polchinski:1987dy} (see \cite{Morinelli:2018pof,Papadopoulos:2024uvi} for further developments and comments), where Polchinski, based on earlier work by Zamolodchikov \cite{Zamolodchikov:1986gt}, was able to prove that scale invariance is necessarily extended to conformal invariance under plausible technical assumptions such as
\begin{itemize}
    \item unitarity (or reflection positivity in the Euclidean case),
    \item a discrete spectrum of operator dimensions,
   \item and existence of the energy momentum tensor and its 2-point functions. 
\end{itemize}
If any of these conditions is given up on, models exhibiting SwCI in $n=2$ may indeed be found \cite{CALLAN197042,PhysRevD.2.753,HULL1986349,Riva:2005gd,Itsios:2021eig,Arutyunov:2015mqj,Naseh:2016maw}. 
\\

Immediately upon the publication of \cite{Polchinski:1987dy}, the question arose whether its insights can be generalized to $n\geq 3$ dimensional field theories\footnote{A brief comment on scale without conformal invariance in $n=1$ can be found in \cite{Nakayama:2013is}. Some of the arguments in \cite{Nakayama:2012ed} may also apply to this case.}. To explain a very complicated topic as briefly as possible (see \cite{Coleman:1970je,Polchinski:1987dy,Nakayama:2013is} for details), conformal invariance requires the (suitably improved) energy momentum tensor $\mathfrak{T}_{mn}$ to be traceless,
\begin{align}
\mathfrak{T}_m^m=0,
\end{align}
while SwCI only requires the trace to be a total derivative,
\begin{align}
\mathfrak{T}_m^m=\partial_m \mathfrak{J}^m. 
\end{align}
Here, $\mathfrak{J}^m$ is called the \textit{virial current}, and the conserved dilatation current reads \cite{Nakayama:2013is}
\begin{align}
\mathfrak{D}_m=x^n \mathfrak{T}_{mn}-\mathfrak{J}_m.
\end{align}
The necessary improvement of the energy momentum tensor is always possible if its trace can be written as
\begin{align}
\mathfrak{T}_m^m&=\partial_m \partial_n \mathfrak{L}^{mn}\text{ for }n\geq 3,
\\
\mathfrak{T}_m^m&=\partial_m \partial^m \mathfrak{L}^{\ \ \,\,}\text{ for }n =2,
\end{align}
for some local operator $\mathfrak{L}^{mn}$ or $\mathfrak{L}$ respectively \cite{Polchinski:1987dy,Nakayama:2013is}. The simplicity of this problem is deceptive, and in the decades since \cite{Polchinski:1987dy} came out, the problem of scale vs.~conformal invariance has to our knowledge not been solved in generality. 

The most significant progress has been made in $n=4$, where over the course of many years increasingly convincing arguments were collected that scale invariance should indeed imply conformal invariance in physical models \cite{MACK1969174,CALLAN197042,PhysRevD.2.753,Zheng:2011bp,Antoniadis:2011gn,Nakayama:2011tk,Nakayama:2011wq,Fortin:2011sz,Fortin:2012hn,Luty:2012ww,Nakayama:2012nd,
Dymarsky:2013pqa,Nakayama:2013is,Yonekura:2014tha,Nakayama:2020bsx}. Many of these arguments are based on or extensions of the $a$-theorem of Komargodski and Schwimmer \cite{Komargodski:2011vj}. See also 
\cite{Dorigoni:2009ra,Nakayama:2013is} for comments on the problem of scale vs. conformal invariance in general dimensions. 
On the other hand, once common conditions of physicality (such as the ones listed above) are relaxed, then indeed models exhibiting SwCI may be found \cite{Iorio:1996ad,Ho:2008nr,Nakayama:2010ye,Nakayama:2012nd,Nakayama:2016xzs,Nakayama:2016ydc,Oz:2018yaz,Mauri:2021ili,Gimenez-Grau:2023lpz,Nakayama:2024jwq,Nakayama:2023wrx}. What about examples of (conventionally) physical models exhibiting SwCI? While we are aware of no such example in $n=4$, there are indeed examples in $n=3$ and $n\geq5$ \cite{Nakayama:2013is}. The by now most famous one is surprisingly simple: Maxwell theory in general dimensions has 
\begin{align}
 \mathfrak{T}_m^m=\frac{4-n}{8}\partial_a \left(A_bF^{ab}\right)
\end{align}
and so exhibits SwCI in $n=3$ and $n\geq 5$  \cite{Jackiw:2011vz,El-Showk:2011xbs}. Something similar happens also in linearized gravity \cite{Farnsworth:2021zgj}. We hence see that the "correct" answer in the left-most column in table \ref{tab::T1} strongly depends on the dimension $n$, as well as on the conditions that an allowable QFT is supposed to satisfy.\footnote{\label{fn5} The example of Maxwell theory is obviously a free theory. While in \cite{El-Showk:2011xbs} it was shown that this theory can be embedded into a non-unitary CFT, a complementary and very interesting result was obtained in \cite{Dymarsky:2015jia}: In any number of dimensions, any sub-sector of a unitary conformal field theory exhibiting only scale without conformal invariance is necessarily a free theory. This might give an indication that specifically in holography, where we often assume strong coupling, SwCI might not be expected to occur.}
\\

Let us now shift our attention to the issue of holographic model building. As said already, the AdS/CFT correspondence \cite{Maldacena:1997re} (or more generally, gauge/gravity duality, see \cite{Ammon:2015wua}) is an important tool to gain insights into stongly coupled field theories that would be hard to obtain from a purely field theory perspective. 
One of its cornerstones is the matching of symmetries in the bulk and boundary. For example, in the prototypical case of AdS$_{n+1}$/CFT$_n$ the isometry group of the $d=n+1$-dimensional AdS-space in the bulk matches with the conformal algebra \eqref{algebras} of the $n$-dimensional CFT\footnote{As is well known and for example pointed out in \cite{Polchinski:1987dy}, CFTs in $n=2$ dimensions are special as the conformal group is extended to the infinite dimensional Virasoro group. In AdS$_3$/CFT$_2$, only the $SO(n=2,2)$ group of the so called M\"obius transformations is actually encoded in the isometry group of the bulk vacuum spacetime, the additional symmetries only arise as \textit{asymptotic} symmetries. 
Throughout this paper, even when $n=2$, we will only be concerned with the proper isometry group and not asymptotic symmetries of the bulk spacetime. As the Virasoro group includes the M\"obius transformations, the proof of \cite{Polchinski:1987dy} that scale invariance implies (infinite dimensional) conformal invariance in $n=2$ of course also trivially means that scale invariance implies invariance under the M\"obius transformations under the same technical assumptions. }. This strategy of symmetry matching has been successfully employed to holographically describe field theories with non-relativistic symmetries \cite{Son:2008ye,Balasubramanian:2008dm,Kachru:2008yh,Bagchi:2009my}, and so the question arises whether SwCI can be studied holographically via bulk metrics with the appropriate isometry algebra. This is where we encounter a first surprise, which was explained nicely in \cite{Nakayama:2010wx} among other places: Suppose that, as common in bottom-up holography, the bulk has only one dimension more than the dual field theory, i.e.~$d=n+1$ in our conventions. Then demanding that the bulk metric contains the $n$-dimensional \Poincare algebra in its isometry algebra necessitates the ansatz
\begin{align}
   ds^2= f(z)dz^2 + g(z) \eta_{ab}dx^adx^b 
\end{align}
with $a,b\in\{0,\cdots,n-1\}$ (in a suitable coordinate system). Now, making the additional requirement that there should be another isometry corresponding to scale transformations fixes the remaining functions $g(z)$ and $f(z)$ (using also residual gauge freedom), and we obtain 
\begin{align}
   ds^2= \frac{1}{z^2}dz^2 + \frac{1}{z^2} \eta_{ab}dx^adx^b. 
   \label{AdS}
\end{align}
But this is just AdS-space, whose isometry group is the full conformal algebra \eqref{algebras}. We hence see that in this naive holographic setup, and extension from scale to conformal invariance takes place automatically as a matter of geometry alone, without the need to appeal to further conditions of physicality such as unitarity on the boundary or energy conditions in the bulk. A holographic study of SwCI must hence find a way to circumvent this limitation. If the bulk gravity is assumed to be Einstein gravity with minimally coupled matter fields, there are roughly two ways to do this\footnote{In \cite{Nakayama:2012sn}, a bulk theory of gravity is considered which is invariant only under foliation preserving diffeomorphisms. The diffeomorphisms corresponding the special conformal transformations are not foliation preserving, so they are automatically invalid as isometries in such a theory, even when the line element formally looks like (\Poincare-patch) AdS \eqref{AdS}. In \cite{Li:2018rgn}, the bulk gravity is considered to be "Einstein-Horndeski gravity", which in essence is Einstein-gravity with a non-minimally coupled shift-invariant scalar field.  
}:
\\

The first option is to allow the bulk metric to have the AdS-form \eqref{AdS}, but require the bulk matter fields to have a coordinate dependence that breaks invariance under special conformal transformations, but not scale transformations. The idea behind this is that a "true" isometry of the bulk configuration should be one that leaves all bulk fields invariant, not just the metric tensor. This ansatz was followed for example in \cite{Nakayama:2009qu,Nakayama:2010wx,Nakayama:2010zz,Nakayama:2011zw,Nakayama:2010ye,Nakayama:2016ydc}.  From our perspective, the drawback of this type of approach is that either backreaction has to be neglected (using the probe limit), or backreaction is taken into account, but then the energy momentum tensor of the bulk matter has to have a cosmological constant-like form $T_{\mu\nu}\propto g_{\mu\nu}$, as otherwise the AdS metric would not be a solution to Einstein's equations any more. Some of the papers utilising this type of ansatz then impose the \textit{strict null energy condition} (strict NEC), which essentially forbids this type of cosmological constant-like backreaction by non-trivial matter fields, see e.g.~\cite{Nakayama:2010wx}. 
\\

The second option is to increase the number of bulk dimensions by adding (warped-) extra dimensions \cite{Nakayama:2009fe,Nakayama:2010zz,Flory:2017mal,Rykala}. It is then possible to find $d$-dimensional bulk metrics whose isometry algebra matches the $n$-dimensional SwCI algebra of \eqref{algebras} with $d\geq n+2$.  This will be exactly our approach in this paper, and we will investigate the validity of our main conjecture (box \ref{B1}) for the specific case of $d=n+2$. 
Out of the existing literature, the papers most closely related to what we intend to do in this publication (apart from the direct precursor works \cite{Flory:2017mal,Rykala}) are thus \cite{Nakayama:2009fe,Nakayama:2010zz}. We would hence like to clearly state how our present work goes beyond these papers, or is complimented by them: 
\begin{itemize}
\item The papers \cite{Nakayama:2009fe,Nakayama:2010zz}, in addition to the NEC, assume a concrete model for the matter fields in the bulk, namely the low energy bosonic actions of String and M-theory. This will restrict the possible forms of the bulk energy momentum tensor (see the problem of geometrisation discussed in section \ref{sec::Synge} and \cite{Krongos:2015fta})\footnote{In addition, the papers \cite{Nakayama:2009fe,Nakayama:2010zz} make the explicit ansatz that the form fields appearing in the bulk are invariant under the same (SwCI) isometries as the spacetime metric. We consider this to be a reasonable assumption whenever specific bulk matter fields are considered (as discussed above and in appendix \ref{sec::scalar}), but for the main results of this manuscript, we exclusively consider the isometries of the metric tensor.}.  We make no assumptions about the bulk matter fields beyond imposing the NEC.


\item  The papers \cite{Nakayama:2009fe,Nakayama:2010zz} assume a certain simplifying gauge condition on the form of the metric, while we assume our metric \eqref{Metric_ansatz} to be as general as possible.

\item One of our central results, discussed in box \ref{B0} and proven in section \ref{sec::Weyl}, is that the bulk Weyl tensor plays a pivotal role in distinguishing spacetimes with SwCI-isometry group from those with a full conformal isometry group. No such importance of the bulk Weyl tensor was commented on in \cite{Nakayama:2009fe,Nakayama:2010zz}. This may be related to the next point:

\item Our results are specific to the case of one warped extra dimension, while the proof of \cite{Nakayama:2009fe,Nakayama:2010zz} does not seem to be dependent on the number of warped extra dimensions. See section \ref{sec::dimensionality} for a detailed discussion of this point. 
\\

\end{itemize}

\noindent
The present work goes beyond the direct precursor works \cite{Flory:2017mal,Rykala} in the following ways:
\begin{itemize}
\item The paper \cite{Flory:2017mal}, as a follow-up work of \cite{Balasubramanian:2013ux}, was primarily concerned with the issue of \textit{discrete scale invariance (DSI)} as opposed to SwCI. The possibility of applying the same kind of warped product ansatz for the bulk metric to the problem of SwCI in holography was only briefly remarked upon in a one-page section at the very end of \cite{Flory:2017mal}. Nevertheless, a detailed study of SwCI along these lines was left to future research at the time, which is now finally provided with this paper.  

\item In \cite{Flory:2017mal}, the potential significance of the Weyl tensor was briefly commented on, but ultimately also left to future investigations. Expanding on \cite{Flory:2017mal}, it was one of the main results of \cite{Rykala} to establish that a non-vanishing bulk Weyl tensor prevents full conformal invariance, hence implying the $\Leftarrow$ direction of the statement in box \ref{B0}. Yet, the method used to prove this result was somewhat of a "brute-force" nature as we explain at the end of section \ref{sec::CIphi0}. Our present paper thus goes beyond these earlier works not only by proving \textit{both} directions of the statement in box \ref{B0} (as well as the additional result in section \ref{sec::det_phi}), but also by presenting a much more elegant and powerful approach to this issue. It should be noted that the $\Rightarrow$ direction of box \ref{B0} is logically indispensable for the proof of our main conjecture to be presented later. 

\item The second main result of \cite{Rykala} was the study of the Killing equations (and their explicit local solutions) for the Killing vector corresponding to scale transformations. This was done mostly in so-called isothermal coordinates on the base space (see the footnote in section \ref{sec::Petrov}) which greatly simplify many expressions and can always be constructed \textit{locally}. 
As the non-trivial topology of the base space is however an important ingredient in our main conjecture in box \ref{B1}, we do not have use for these specific results in this paper.  

\item The statement of box \ref{B1} was also mentioned in \cite{Rykala} as a conjecture, but no proof was provided, not even for simplified special cases. We hence go far beyond \cite{Rykala} with our results in section \ref{sec::proofs} and appendices \ref{sec::1Dfiber} and \ref{sec::higherDfiber}. In addition, the reformulation of the null energy condition in section \ref{sec::NEC} which is the groundwork for these proofs (as well as the weak energy condition in appendix \ref{sec::WEC}) is a new result of our work. 


\end{itemize}

\noindent
We will now proceed to give a detailed discussion of the geometry of the warped-product ansatz for the bulk metric in the next section.

\section{Warped Product Ansatz and Killing Algebra}
\label{sec::WPA}

\subsection{Ansatz and Notation}
\label{sec::AnsatzAndNotation}

The ansatz made in \cite{Balasubramanian:2013ux,Flory:2017mal,Rykala} was to realise bulk metrics with spacetime isometries such as discrete scale invariance (DSI) or scale without conformal invariance (SwCI) via the introduction of a warped product ansatz of the form 
\begin{align}
ds^2=e^{2C(w,\theta)}\left[e^{\frac{2w}{L}}(-dt^2+d\vec{x}^2)+dw^2\right]+e^{2B(w,\theta)}(d\theta+A(w,\theta)dw)^2
\label{Balasubramanian_Metric}
\end{align}
where the expression in the square brackets corresponds to the line element of AdS, and $\theta$ is the warped extra dimension which is assumed to be compact. However, as the AdS-space is in itself already a warped product between the Minkowski space and the $w$-direction, we find it much more convenient to work with an ansatz of the form \cite{Rykala} 
\begin{align}
ds^2=g_{\mu\nu}dx^\mu dx^\nu = S(w,\theta) \underbrace{\eta_{ab} dx^a dx^b}_{\text{fiber}} + \underbrace{h_{ij}dy^i dy^j}_{\text{base space}}
\end{align}
which is a warped product between Minkowski space (with coordinates $x^a\in\{t,\vec{x}\}$) and a two-dimensional base space with coordinates $y^i\in\{w,\theta\}$ and the topology of a cylinder ($w\in\mathbb{R},\ \theta\sim\theta+2\pi$). For the metric to have correct signature, we demand $S>0$ everywhere, and throughout the paper we will use $d$ for the dimension of the entire bulk, and $n$ for the dimension of the Minkowski fiber ($d=n+2$ for a 2-dimensional base space). This ansatz is the most general one consistent with an isometry algebra of the form of the \Poincare algebra in \eqref{algebras}. Specifically, the metric components are independent of the coordinates $\{t,\vec{x}\}$ because of the invariance under translations, and there are no terms of the form $g_{ai}dx^a dy^i$ due to the invariance under boosts and rotations. 
\\

We begin by introducing some nomenclature and useful notation. As already indicated in \eqref{Metric_ansatz}, the Minkowski space in this ansatz is referred to as the "fiber", while the space spanned by the extra dimensions $w$ and $\theta$ is called the "base space" \cite{chen2017differential}. Throughout the paper, it will prove useful to rephrase geometrical problems concerning the bulk geometry in the language of the 2-dimensional geometry of the base space. We will thus, as in \eqref{Metric_ansatz}, use Latin indices $i,j,...$ for the base space, Greek indices $\mu,\nu,...$ for all bulk coordinates, and Latin indices $a,b,...$ in the rare instances where we need the coordinates in the fiber. We will also mark quantities defined in the geometry of the base space with a bar, e.g.~the Christoffel symbols, Ricci tensor and Lie derivative of the 2-dimensional metric $h_{ij}$ are $\bar{\Gamma}^i_{jk}$, $\bar{R}_{ij}$, and $\bar{\mL}$ respectively, while the 4-dimensional Christoffel symbols, Ricci tensor and Lie derivative of the whole bulk metric $g_{\mu\nu}$ are $\Gamma^\alpha_{\mu\nu}$, $R_{\mu\nu}$ and $\mL$. There are of course well-known relations for warped product spacetimes that express the overall curvature tensors in terms of the curvature tensors of base space and fiber (see e.g.~\cite{chen2017differential,Nozawa:2008rjk}\footnote{See also the Mathematica files accompanying this publication at \cite{UJ/J9M1LS_2025}.}), some of which we will make use of later when they become relevant. 
For derivatives of $S(w,\theta)$, we write
\begin{align}
(\nablaII S)^2&\equiv \nablaII_i S \nablaII^i S,
\\
\boxII{S}&\equiv\nablaII^i\nablaII_i S,
\end{align}
and introduce the tensor
\begin{align}
    q_k\equiv\nablaII_k \log{\sqrt{S} }=\frac{\nablaII_k\sqrt{S}}{\sqrt{S}}.
    \label{DefOfq}
\end{align}
It is then easy to show the following equalities
\begin{align}
\nablaII_i q_j &= \nablaII_j q_i\\
    \frac{(\nablaII \sqrt{S})^2}{S}  &= q_k q^k  \equiv q^2  \\
       \frac{\nablaII_i \nablaII_j\sqrt{S} }{\sqrt{S}} &= (\nablaII_i q_j + q_i q_j)  
       \\ \frac{\boxII \sqrt{S}}{\sqrt{S}} &= \nablaII_k q^k + q_k q^k \equiv \nablaII q + q^2
  \end{align}
which will be useful later. For the sake of concreteness, we will from now on only focus on the case of $d=4$ bulk dimensions ($\vec{x}=x$), so that the putative holographic dual would be an $n=2$ dimensional field theory. We will come back to the issue of general bulk dimensions later in section \ref{sec::dimensionality} and appendix \ref{sec::higherDfiber}. 
For future reference, we will also show explicitly how our results and ideas work out for the simplest possible case, AdS$_3\times$S$^1$, at the end of this and some of the later subsections:

\ADS{
The ansatz \eqref{Metric_ansatz} simplifies to \ADSxS 
for the choices
\begin{align}
h_{ij}&=\delta_{ij} \Rightarrow \bar{R}=0,
\label{AdSS1R}
\\
S&=e^{\frac{2w}{L}} \Rightarrow q_i=\left(
\begin{array}{cc}
 \frac{1}{L}\ ,& 0\\
\end{array}
\right),\ \nablaII q=0,\ q^2 = \frac{1}{L^2}.
\label{AdSS1q}
\end{align}
}

\subsection{Killing Algebra}
\label{sec::KillingAlgebra}

We are interested in those spacetimes of the form \eqref{Metric_ansatz} whose isometry group corresponds to Scale without Conformal Invariance (SwCI), i.e.~whose Killing vector fields form the SwCI (sub-)algebra of \eqref{algebras} where the brackets are the regular Lie-brackets 
\begin{align}
\left[\mV_{1},\mV_{2}\right]^{\mu}\equiv\mV_1^\rho\partial_{\rho}\mV_2^\mu-\mV_2^\rho\partial_{\rho}\mV_1^\mu.
\label{bracket}
\end{align}
As is well known, the Lie-bracket of two Killing fields $\mV_1$, $\mV_2$ will itself be a Killing field. 
It is quite clear that \eqref{Metric_ansatz} allows for Killing vectors generating the full \Poincare invariance, as due to the warped product construction, the Killing fields of the fiber space can be lifted to Killing fields in the full geometry. 
As in \cite{Flory:2017mal,Rykala}, one can now show that if a Killing vector corresponding to scale transformations exists on this spacetime background, the Lie-algebra \eqref{algebras} alone restricts its possible form to 
\begin{align}
D^\mu=\left(\begin{array}{c}
-t\\
-x \\
J^w(w, \theta) \\
J^\theta(w, \theta)
\end{array}\right).
\label{D-type}
\end{align}
As we see, the 4-dimensional vector $D^\mu$ is fully determined by a 2-dimensional vector $J^i$ living in the geometry of the base space. Similarly, for Killing vectors potentially generating special conformal transformations, we find \cite{Flory:2017mal,Rykala}
\begin{align}
(K_0)^\mu=\left(\begin{array}{c}
\mathcal{K}(w, \theta)+t^2+x^2\\
2 t x \\
-2 t J^w(w, \theta) \\
-2 t J^\theta(w, \theta)
\end{array}\right),\ \ 
(K_1)^\mu=\left(\begin{array}{c}
-2 t x \\
\mathcal{K}(w, \theta)-t^2-x^2\\
2 x J^w(w, \theta) \\
2 x J^\theta(w, \theta)
\end{array}\right),
\label{Ktypes}
\end{align}
with the additional constraint \cite{Flory:2017mal,Rykala}
\begin{align}
2\mathcal{K}(w, \theta) + J^\theta(w, \theta)\partial_{\theta}\mathcal{K}(w, \theta) + J^w(w, \theta)\partial_{w}\mathcal{K}(w, \theta) = 0.
\label{mKequationLong}
\end{align}
Whether any of the vectors in \eqref{D-type} and \eqref{Ktypes} are actual Killing vectors however depends of course on whether they satisfy the Killing equation
\begin{align}
\mL_\mV g_{\mu\nu}=\nabla_{\mu}\mV_{\nu}+\nabla_{\nu}\mV_{\mu}=0, 
\label{Killing}
\end{align}
where we have used $\mL$ for the Lie derivative of a tensor with respect to a vector field. Plugging the ansatz \eqref{D-type} into \eqref{Killing}, we obtain three cases, depending on whether the indices $\mu,\nu$ are chosen from the fiber or base-space coordinates \cite{Rykala}:
\begin{align}
\mu=a, \nu=j:&\nonumber
\\
&0=0\ \text{  trivially},
\\
\mu=a, \nu=b:&\nonumber
\\ 
&0=\left(-2 S + \bar{\mL}_J S\right)\eta_{ab},
\label{Sequation}
\\
\mu=i, \nu=j:&\nonumber
\\ 
&0= \nablaII_i J_j + \nablaII_j J_i =\bar{\mL}_J h_{ij}.
\label{Fequation}
\end{align}
Equation \eqref{Sequation} is a differential equation for the function $S$, while \eqref{Fequation} is nothing but the Killing equation in the 2-dimensional base space. To summarise, in order for a metric of the form \eqref{Metric_ansatz} to include scale invariance in its isometry group, the base space geometry $h_{ij}$ needs to have a Killing vector $J^i$, and the function $S$ needs to solve the differential equation \eqref{Sequation}. It is however important to understand that this equation will have an infinite number of solutions, as if $S$ is a solution to \eqref{Sequation} and $G(w,\theta)$ is a function such that $\bar{\mL}_J G=0$, then $f(G)\cdot S$ will also be a solution to \eqref{Sequation} for any function $f$ \cite{Rykala}.
\\

Using equations \eqref{Sequation} and \eqref{DefOfq}, we can already show
\begin{align}
J^i q_i=1.
\label{Jq1}
\end{align}
As the base space has only two dimensions, this means that $q^i$ can be decomposed into a component parallel to $J^i$ and one component perpendicular to it. A useful ansatz will hence later be
\begin{align}
    q_i = \frac{J_i}{J^2} + \nablaII_i \log{Q},
    \label{qQansatz}
\end{align}
where $Q$ is a scalar function s.t.~$\bar{\mL}_J Q= 0$ and it is evident that $J^i q_i = 1$ as required. From this ansatz, we see that the divergence of $q$ is solely determined by $Q$, as
\begin{align}
   \nablaII_i q^i 
   =\boxII \log{Q}.
    \label{DivqQ}
\end{align}
The term $\nablaII_i  \frac{J^i}{J^2}$ vanishes as $J^i$ is a Killing vector. 
\\

Of course, as we are only interested in scale \textit{without} conformal invariance, we need to search for metrics of the form \eqref{Metric_ansatz} where $D^\mu$ (equation \eqref{D-type}) is a Killing vector (i.e.~\eqref{Sequation} and \eqref{Fequation} can be satisfied), but where the potential generators of special conformal transformations \eqref{Ktypes} do \textit{not} satisfy the Killing equation. We will come back to this important issue in section \ref{sec::Weyl}, but first we will collect some other useful information in the rest of this section.

\ADS{
\ADSxS contains Killing vectors forming the full conformal algebra in addition to translation invariance along the compact extra dimension $\theta$. We explicitly find
\begin{align}
J^i&=\left(
\begin{array}{c}
 L \\
 0  \\
\end{array}
\right)
\Rightarrow 
J^2=L^2,\ 
D^\mu=\left(\begin{array}{c}
-t\\
-x \\
L \\
0
\end{array}\right),\ 
    q_i = \frac{J_i}{J^2}.\hfill
\label{JforAdSS1}
\end{align}
}

\subsection{Topology of the Base Space}
\label{sec::Topology}

Equation \eqref{Fequation} shows that the metric $h_{ij}$ must have at least one Killing vector, $J^i$. The following is a well-known fact:

\BigBox{\begin{defn}\label{B2}Classification of 2d metrics\end{defn}}
{
In 2 dimensions, a metric can have either
\begin{itemize}
\item no Killing vectors,
\item one Killing vector,
\item or three linearly independent Killing vectors
\end{itemize}
in any infinitesimal neighbourhood. The latter case corresponds to a maximally symmetric metric with constant curvature $\bar{R}$.
}

The case of exactly two linearly independent Killing vector fields does not exist, because the Ricci scalar $\bar{R}$ has to be constant along flows of the Killing vector fields. But in two dimensions, two linearly independent Killing vector fields would imply the curvature to be constant in all directions, which in 2 dimensions is enough to imply that the spacetime is locally maximally symmetric. 
There are some results in the mathematical literature concerning the classification of 2d-metrics admitting exactly one Killing field locally, see for example \cite{Bryant_2007,Matveev_2012}. However, the big issue that we are facing is the issue of topology.
\\

Consider for instance the case of vanishing $\bar{R}$ everywhere. That means locally the metric $h_{ij}$ of the base space is just the metric of the flat plane, with three linearly independent Killing vector fields in any small enough neighbourhood. However, as stated in Box \ref{B1}, we demand the base space to have the topology of a cylinder. Folding up the flat space to a cylinder preserves two linearly independent Killing vectors globally, but one of the Killing vector fields can not be extended globally in a single valued manner any more. 
We will from now on only consider those Killing vector fields to be valid which can be extended globally over the whole base space in a single valued manner\footnote{This seems to be a common attitude in the GR community, for instance the BTZ black hole in 3d is well known to be constructed by folding up the (maximally symmetric) AdS$_3$ space, yet it is commonly described as having only two Killing vectors \cite{Banados:1992gq}.}.  
\\

\begin{figure}
    \centering
    \includegraphics[width=0.95\linewidth]{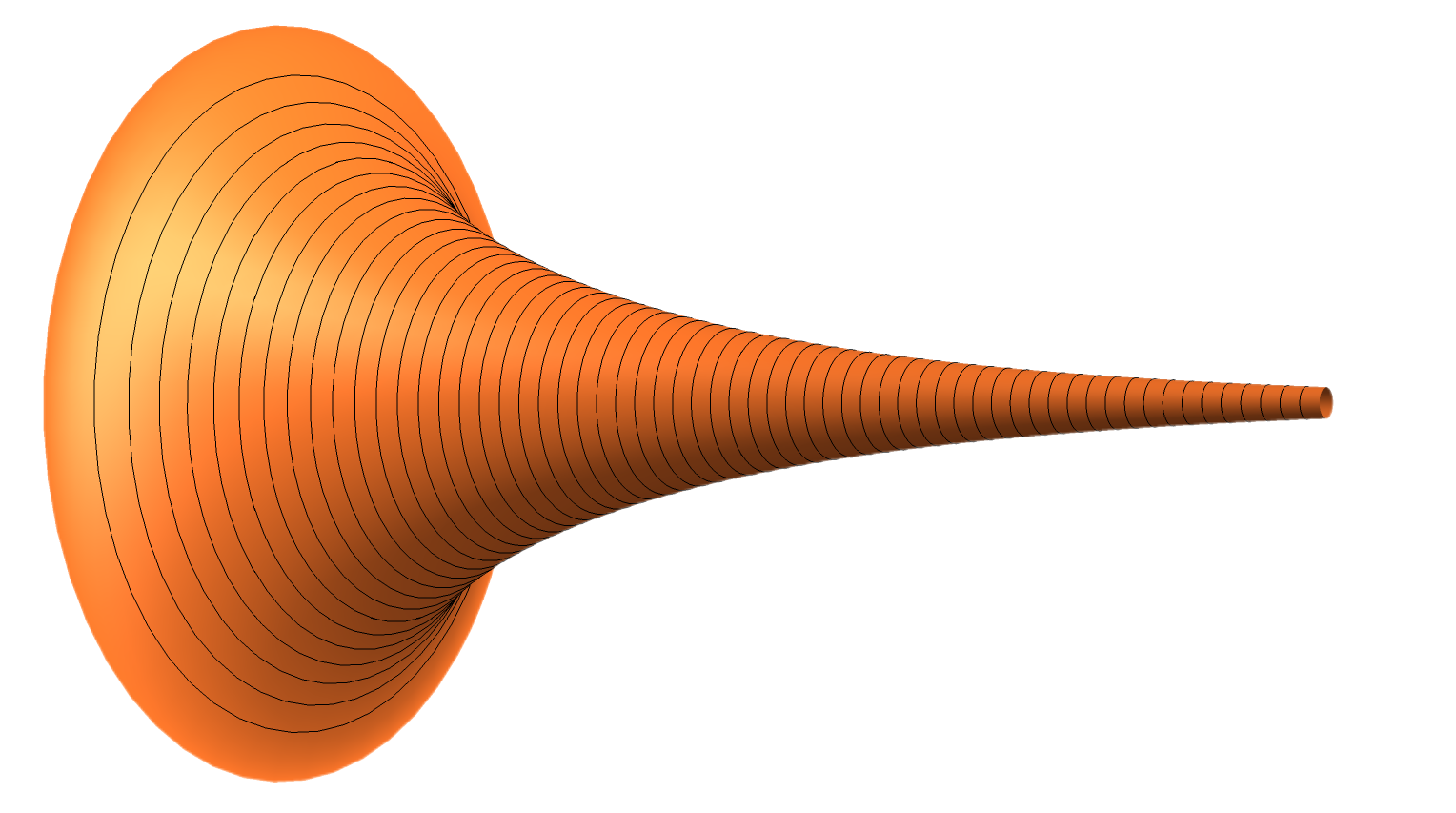}
    \caption{Embedding of the pseudosphere into $\mathbb{R}^3$ as a surface of revolution. The black lines are integral lines of the only globally defined Killing vector field.  }
    \label{fig::ps}
\end{figure}

There is an additional restriction arising from the non-trivial topology that we have to consider. 
To give an example, following the classification provided in Box \ref{B2} and the discussion above, one might decide to investigate the case where the base space is given by a pseudosphere \cite{Beltrami1868, krivoshapko2015encyclopedia}, see figure \ref{fig::ps}. This is a 2-dimensional space with constant negative curvature, thus locally isometric to the hyperbolic plane, but with the topology of a cylinder. 
It can be constructed by a nontrivial identification of the 2d hyperbolic space, not unlike the BTZ black hole in one dimension higher \cite{Banados:1992gq}. Out of the three linearly independent solutions of the Killing equation that can be constructed in any simply connected neighbourhood, only one can be extended in a single valued manner to the whole manifold. Furthermore, the pseudosphere can be embedded into $\mathbb{R}^3$ as a surface of revolution, and so the global Killing vector field on it is the one that describes the rotational invariance. Consequently, its integral lines are circles. But equation \eqref{Sequation} generally implies that $S(w,\theta)$ will be a monotonously increasing function along the flow of the Killing vector field $J^i$. That means that in our spacetime construction, while the base space must have the topology of a cylinder, the integral lines of $J^i$ must not be compactified to a circle for $S$ to be a single-valued monotonously growing function. 
\\

Such models can hence not serve as a possible base spaces in our constructions. To summarise, we have to demand the following: 
\BigBox{
\begin{defn}\label{B3}Topology of the base space\end{defn}
}
{
The base space in \eqref{Metric_ansatz} must have the following properties:
\begin{itemize}
\item It has the topology of a cylinder (see Box \ref{B1}).

\item It allows for (at least) one globally defined (i.e.~single valued) Killing vector field $J^i$ (see \eqref{Fequation}).

\item The integral lines of $J^i$ are not compact, i.e.~they have the topology of $\mathbb{R}$, not $S^1$.

\end{itemize}
}

It is a very easy exercise to generate such manifolds via isometric embeddings.  If $\mathbb{R}^3$ is spanned by a cylindrical coordinate system with coordinates $w\in \mathbb{R},\theta\in [0,2\pi [,\rho\in \mathbb{R}^{+}$, then a "flat" cylinder (i.e.~with flat induced metric) is described by the parametric equation $\rho(w,\theta)=const$. If instead we write $\rho(w,\theta)=f(a\cdot w + \theta)$ with some $a>0$ and some periodic but positive function $f$, we obtain an embedding for a surface with the rough shape of a corkscrew. This surface will have the topology of a cylinder and its induced metric will have one Killing vector field, where the integral lines of this vector field are not compact. Instead, they wind around the surface in a spiral manner. 
\\

There is now one final issue relating to topology that needs to be considered. Essentially, we want to know whether and how we can apply results like the Gauss-Bonnet theorem to the base space as this will be a powerful tool later in this paper. As the base space has infinite extent in one direction, its volume integral is infinite and hence expressions like $\int \bar{R} \sqrt{h}dy^2$ may be ill-defined. However, we point out that if a space has an isometry, then it can be compactified via a quotient construction that identifies points along the integral lines of the Killing field. This is exactly what is done for the examples of the Pseudosphere \cite{Beltrami1868, krivoshapko2015encyclopedia} and the BTZ black hole \cite{Banados:1992gq} mentioned above - the folding process is always related to a Killing vector field with non-compact integral lines in the covering space. The same can be done for the models of the base space that we are considering. So it is important to understand that while for the final model \eqref{Metric_ansatz} the base space should not be compactified along the $w$-direction (to ensure the conditions of box \ref{B3}), it is in principle \textit{compactifyable} to a space with the topology of a torus. Clearly, after such an additional (fictitious) compactification, $S$ will not be defined in a single valued manner anymore, but all geometrical quantities that are invariant under Lie-derivatives with respect to the Killing vector $J^i$ will be. This includes $\bar{R}$ and all other curvature tensors constructed from $h_{ij}$, but notably also the tensor $q_i$ \eqref{DefOfq} and all geometric quantities derived from it via covariant derivatives or contractions:
\begin{align}
    \bar{\mL}_J q_i =  \bar{\mL}_J\nablaII_i \log{\sqrt{S} }
    =  \nablaII_i \bar{\mL}_J\log{\sqrt{S} }
          =  \nablaII_i 1 =0
\end{align}
using commutation relations for Lie- and covariant derivatives \cite{Tanski2020}.

\ADS{
As we already showed in \eqref{AdSS1R}, \ADSxS corresponds to the case $\bar{R}=0$ everywhere, i.e.~where the base space is a flat cylinder. However, because solutions to \eqref{Sequation} are not unique, the converse is not true: There are metrics of the form \eqref{Metric_ansatz} distinct from \ADSxS for which the base space is the flat cylinder, but which may exhibit Scale without Conformal Invariance as we will show below. 
}

\subsection{Curvature and the Gauss-Bonnet theorem}
\label{sec::GBandCurvature}

As a consequence of the discussion in the preceding section, we can formally treat the base space as if it was further compactified from a cylinder to a torus in certain contexts. This means we can now straightforwardly apply the Gauss-Bonnet theorem \cite{Bonnet1848, Renteln_2013}\footnote{
An alternative but ultimately equivalent approach would have been to keep treating the base space as an infinite cylinder, and introduce two cuts to obtain a finite piece. There is of course a version of the Gauss-Bonnet theorem for manifolds with boundaries that can be used in this case \cite{doCarmo:1976}. If the second cut is obtained by transporting the first one along the flow of the Killing vector field on the base space, then one can argue that the extra terms, which depend on the geodesic curvatures of the two cuts, cancel each other exactly, leaving only \eqref{GBtheorem} as $\chi_{\text{Torus}}=\chi_{\text{Annulus}}$.
}
\begin{align}
    \int\frac{\bar{R}}{2} \sqrt{h}d^2y
    = 2\pi\chi_{\text{Torus}}=0.
    \label{GBtheorem}
\end{align}
This result is of great importance for us: It shows that the curvature $\bar{R}$ of the base space is either zero everywhere or both negative somewhere and positive somewhere else. As due to the isometry $\bar{R}$ is constant along the integral lines of the Killing vector $J^i$ (and these integral lines are not compact in the true base space), this essentially means that if we circumnavigate the cylindrical base space on a closed path, $\bar{R}$ will be either zero everywhere or both negative and positive somewhere on this path. 
Referring back to box \ref{B2}, this shows that we are only interested in those metrics for the base space with either exactly one Killing vector or $\bar{R}=0$ everywhere. Metrics with constant $\bar{R}\neq 0$ are not suitable for our model building\footnote{This shows once again that the negatively curved pseudosphere (which violates point 3 in box \ref{B3}) can not serve as a base space. This is related to Bochner's theorem \cite{Bochner48}, which states that on a closed Riemannian manifold of negative Ricci curvature, there is no nontrivial Killing vector field.}.
\\

Analogously to \eqref{GBtheorem}, we can treat any total derivative that is well-behaved under the fictitious compactification of the base space from cylinder to torus. As argued at the end of the previous subsection, this will include derivatives of $q^i$:
\begin{align}
    \int\nablaII_i q^i \sqrt{h}d^2y  =0,
\end{align}
i.e.~$\nablaII q$ is either zero everywhere or both negative somewhere and positive somewhere else when circumnavigating the base space on a closed path. 
In fact, we will now explicitly show that the Ricci curvature $\bar{R}$ of the base space is a total derivative of a remarkably simple and useful quantity. 
To do so, we remind ourselves that the base space has to have (at least) one Killing vector, and in this case it should always be possible to make a local choice of coordinates such that the metric is independent of one of the coordinates. So let us assume we have coordinates $w,\theta$, such that 
\begin{align}
&h_{ij}=
\left(
\begin{array}{cc}
 h_{ww}(\theta) & h_{w\theta}(\theta) \\
 h_{w\theta}(\theta) & h_{\theta\theta}(\theta)  \\
\end{array}\right)
\label{MetricWithKillingAnsatz}
\end{align}
and the Killing vector reads 
\begin{align}
&J^i=
\left(
\begin{array}{c}
1 \\
 0  \\
\end{array}\right) \text{   with   } J^2=J^iJ_i=h_{ww}(\theta).
\label{KillingAnsatz}
\end{align}
A brute force calculation in this coordinate system then reveals the identity
\begin{align}
\sqrt{h}\bar{R}=-\partial_\theta \left(\frac{h_{ww}'(\theta)}{\sqrt{h}}\right),
\label{Rfromf0}
\end{align}
which can be covariantised to the form
\begin{align}
\bar{R}=-\frac{1}{\sqrt{h}}\partial_i \left(\sqrt{h}h^{ij}\partial_j \log(J^2)\right)=-\boxII \log(J^2).
\label{RfromJ2}
\end{align}
holding in any coordinate system. 
\\

This equation may seem surprising at first, so we subject it to some consistency checks and further scrutiny. As stated above, it makes sense in our case that $\bar{R}$ would be a total derivative because of \eqref{GBtheorem}. But what about other compact spaces such as the sphere? In that case, \eqref{RfromJ2} still holds but there will be points where $J^2=0$ and hence $\log(J^2)$ diverges\footnote{
This is related to the Hairy ball \cite{H1885, Brouwer1912, Renteln_2013} and  Poincar\'{e}-Hopf \cite{H1885, Hopf1927, Renteln_2013} theorems.
}. Because of these singularities, \eqref{RfromJ2} can not be used to argue that $\int \bar{R}\sqrt{h}dy^2$ has to vanish. Note that in our case, we require $S>0$ everywhere to ensure that the metric \eqref{Metric_ansatz} has the correct signature, and using \eqref{Sequation} this means that there must not be any point in the base-space (which has a non-degenerate Euclidean metric $h_{ij}$) where $J^i$ becomes the zero vector. Equation \eqref{RfromJ2} is hence always well-defined in the cases of interest to us.  Additionally, if $J^i$ is a Killing vector, then clearly so is $\lambda\cdot J^i$ with some constant $\lambda\neq0$. But $\bar{R}$ can't depend on such an overall scaling choice. This is indeed the case with \eqref{RfromJ2}, because
\begin{align}
\boxII \log(\lambda^2J^2)=\boxII \log(J^2)+\boxII\log{\lambda^2}=\boxII \log(J^2).
\end{align}
Of course, not every metric in 2 dimensions has a Killing vector, but there will always be conformal Killing vectors at least locally because of the conformal flatness of the metric. In appendix \ref{sec::conformal_killing_R}, we will show that \eqref{RfromJ2} actually also holds if $J$ is only a conformal Killing vector. 
\\

Lastly, it is clear from our result \eqref{RfromJ2} that $J^2=const. \Rightarrow \bar{R}=0$. But is the converse also true?
We assume that in appropriate coordinates, a metric of the form \eqref{MetricWithKillingAnsatz}, with a Killing vector like \eqref{KillingAnsatz}, covers the \textit{whole} base space. Because of the cylindrical topology of the base space, we thus assume the periodic identification $\theta\sim \theta+2\pi$ and all metric components and geometrical quantities to be periodic functions of $\theta$. In fact, $J^2$ is a function of $\theta$ only, because of \eqref{KillingAnsatz} but also more elegantly because for a Killing vector we have 
\begin{align}
    \bar{\mL}_JJ^2=0. 
    \label{LieJ2}
\end{align}
The same holds true for \textit{any} scalar curvature quantity defined from the geometry of the base space alone (such as $\bar{R}$, but not $S$ for example). Through an explicit calculation, we then find 
\begin{align}
\bar{R}\equiv0&\Rightarrow
0=\boxII \log(J^2)=\frac{1}{\sqrt{h}}\partial_\theta h^{\theta\theta}\sqrt{h}\,\partial_\theta\log(J^2)
\\
&\Rightarrow const.\equiv h^{\theta\theta}\sqrt{h}\,\partial_\theta\log(J^2).
\label{const}
\end{align}
For a non-degenerate Euclidean metric $h^{\theta\theta}\sqrt{h}>0$, but $\partial_\theta\log(J^2)$ is a total derivative of a periodic function, and so is either zero everywhere or switches sign somewhere. Hence \eqref{const} is only possible if 
\begin{align}
 \partial_y\log(J^2)=0 \Rightarrow J^2=const.   
\end{align}
Consequently, at least in the cases of relevance to us, we find
\begin{align}
\bar{R}=0 \Leftrightarrow J^2=const.
\label{boxLogJ20}
\end{align}
\ADS{
This is clearly consistent with $\bar{R}=0$ (equation \eqref{AdSS1R}) and \eqref{JforAdSS1} because 
\begin{align}
J^2=L^2.    
\label{J2=L2}
\end{align}
}

\subsection{Null Energy Condition}
\label{sec::NEC}

As we indicated in our main conjecture in box \ref{B1}, we will use the \textit{null energy condition} (NEC) to judge whether a bulk metric is physically valid or not. The NEC states that for any null vector $n^\mu$, we have \cite{Curiel:2014zba} 
\begin{align}
0\leq T_{\mu\nu}n^\mu n^\nu.
\label{NEC}
\end{align}
As we are dealing with null vectors, we can replace $T_{\mu\nu}$ with $G_{\mu\nu}$ or $R_{\mu\nu}$ when assuming Einstein's equations. It will be beneficial to write this in a form adapted to our warped product ansatz explained earlier in section \ref{sec::AnsatzAndNotation}. Specifically, we decompose the null vector as
\begin{align}
&n^{\mu}=
\left(
\begin{array}{c}
 \hat{n}^a \\
 \bar{n}^i  \\
\end{array}
\right),
\label{null-vector}
\end{align}
where $\bar{n}^i$ is a vector in the base space and $\hat{n}^a$ is a vector in the (tangent space of the) fiber. This implies
\begin{align}
0\equiv n^\mu g_{\mu\nu} n^\nu = S \hat{n}_a\hat{n}^a+\bar{n}^i h_{ij}\bar{n}^j \Rightarrow \hat{n}_a\hat{n}^a = -\frac{\bar{n}^i h_{ij}\bar{n}^j}{S}.
\end{align}
Next, we use that in the warped product ansatz \eqref{Metric_ansatz}, one finds \cite{Nozawa:2008rjk} 
\begin{align}
R_{\mu\nu}=
\left(
\begin{array}{cc}
 X \eta_{ab} & 0 \\
 0 & r_{ij}  \\
\end{array}
\right)
\label{RmunuForm}
\end{align}
with, for the case of a two-dimensional fiber on which we are focusing at the moment,
\begin{align}
    X&= -[\sqrt{S} \boxII\sqrt{S}+ (\nablaII \sqrt{S})^2]
    \\
    &=-S[\nablaII q + 2 q^2], 
    \label{Xdef}
\end{align}
and
\begin{align}
r_{ij} &= \frac{\bar{R}}{2} h_{ij} - \frac{2}{\sqrt{S}}\nablaII_i\nablaII_j\sqrt{S}
\\
    &=\frac{\bar{R}}{2} h_{ij} -2
    (\nablaII_i q_j + q_i q_j)  .
\end{align} 
The NEC hence simplifies to
\begin{align}
0&\leq T_{\mu\nu}n^\mu n^\nu 
\\
&= R_{\mu\nu}n^\mu n^\nu
\\
&= X \hat{n}_a\hat{n}^a + \bar{n}^i r_{ij}\bar{n}^j
\\
&= \left(r_{ij}-\frac{X}{S}h_{ij}\right)\bar{n}^i\bar{n}^j\ \ \forall \bar{n}^j.
\label{NECforn}
\end{align}
So, the question becomes whether 
\begin{align}
t_{ij}&\equiv r_{ij}-\frac{X}{S}h_{ij}
\\
&=\left[\frac{\bar{R}}{2}+\frac{\boxII\sqrt{S}}{\sqrt{S}}+\frac{(\nablaII \sqrt{S})^2}{S} \right]h_{ij}- \frac{2}{\sqrt{S}}\nablaII_i\nablaII_j\sqrt{S}
\\
&=\left[\frac{\bar{R}}{2}+\nablaII q + 2q^2 \right]h_{ij}- 2\nablaII_i q_j -2 q_i q_j
\label{tij}
\end{align}
is positive semi-definite or not for a given $S$ and $h_{ij}$ and at any possible location on the base space manifold. 
\\

Of course, the base space is Euclidean with a positive definite metric so locally one can always choose a coordinate system where $h_{ij}=\delta_{ij}$, and $t_{ij}$ can be treated like a regular $2\times2$ matrix. Such a matrix can be characterized by its trace and determinant:
\begin{align}
\text{tr}[t_{ij}]\geq 0 \text{ and } \det[t_{ij}]\geq 0 &\Leftrightarrow t_{ij}\text{  is positive semi-definite}
\Leftrightarrow \text{NEC is satisfied,}
\label{NEC2d}
\\
\text{tr}[t_{ij}]\leq 0  \text{ and } \det[t_{ij}]\geq 0 &\Leftrightarrow t_{ij}\text{  is negative semi-definite,}
\\
\det[t_{ij}]<0 
&\Leftrightarrow t_{ij}\text{  is indefinite.}
\end{align}
The trace is easy to determine:
\begin{align}
\text{tr}[t_{ij}]=t^i_j=\bar{R}+\frac{2(\nablaII \sqrt{S})^2}{S} =\bar{R}+2 q^2.
\end{align}
To solve for the determinant, we make use of two identities that are valid for $2\times2$-matrices:
    \begin{align}
       \det\left[A+B\right] &= \det[A]+\det[B]+\Tr[A]\Tr[B]-\Tr[AB],
       \\
      \det[B]&=\frac{1}{2}\left(\Tr[B]^2-\Tr[B^2]\right).
    \end{align}
Thus, we obtain
\begin{align}
\label{DETtij}
\det[t_{ij}]&=\frac{1}{4}\left(\bar{R}+\frac{2(\nablaII \sqrt{S})^2}{S}\right)^2+\frac{1}{S}\left(\boxII \sqrt{S}\right)^2-\frac{2}{S} (\nablaII_i\nablaII_j \sqrt{S})(\nablaII^i\nablaII^j\sqrt{S})
\\
&=\frac{1}{4}\left(\bar{R}+2q^2\right)^2+ (\nablaII q + q^2)^2 - 2 (\nablaII_i q_j + q_i q_j)(\nablaII^i q^j + q^i q^j).
\label{DETtijq}
\end{align}

\ADS{
With what we already know, it is easy to find that
\begin{align}
t_{ij}= 2q^2 h_{ij}-2 q_i q_j,\ \text{tr}[t_{ij}]=\frac{2}{L^2},\ \det[t_{ij}]=0.
\end{align}
Clearly, the NEC is satisfied everywhere.
}

In appendix \ref{sec::WEC}, we will briefly discuss the weak energy condition (WEC) in warped product spacetimes of the type that we are investigating, however we will show that it doesn't provide any additional information or constraints on the spacetime compared to the NEC. In appendix \ref{sec::higherDfiber}, we will discuss how the expression for $t_{ij}$ depends on the number of dimensions of the fiber.

\section{Role of the Weyl tensor}
\label{sec::Weyl}

\subsection{Form of the Weyl tensor}
\label{sec::Weyl_form}

In the earlier paper \cite{Flory:2017mal} it was already noticed that, but not fully understood why, the bulk Weyl tensor plays a pivotal role in determining whether the bulk isometry group can be extended to full conformal invariance. This may seem natural at first, after all it is well known \cite{Stephani:2003tm} that the Weyl tensor is invariant under conformal rescalings of the metric, and vanishes for conformally flat metrics. See e.g.~\cite{Iorio:1996ad} for information on the relations between scale, Weyl, and conformal invariance. 
However, when we talk about scale or conformal invariance in the present work, we always mean scale and special conformal transformations in a putative holographically dual boundary theory, not scale or conformal transformations in the bulk. Hence, upon further consideration, it is not obvious to us at all why the Weyl tensor should have been expected to be so important. 
As we will show in this section, the vanishing or not-vanishing of the Weyl tensor allows us to distinguish between those metrics in which scale invariance is extended to full conformal invariance, and those metrics that correspond to genuine scale- without conformal invariance. 
\\

To begin, we note that for a generic metric, the Weyl tensor
\begin{align}
C_{\alpha\beta\gamma\delta}\equiv R_{\alpha\beta\gamma\delta}
-\frac{1}{2}\left(g_{\alpha[\gamma }R_{\delta ]\beta}-g_{\beta[\gamma }R_{\delta ]\alpha}\right)
+\frac{1}{3}Rg_{\alpha[\gamma }g_{\delta ]\beta}
\label{WeylAndReducedWeyl}
\end{align}
depends on the metric and up to second derivatives of it, and consequently can be quite complicated. For a metric of the warped product form \eqref{Metric_ansatz} however, the Weyl tensor can be written in a form
\begin{align}
C_{\alpha\beta\gamma\delta}=\phi\, \tilde{C}_{\alpha\beta\gamma\delta},\ \         C_{\mu \nu \alpha \beta} C^{\mu \nu \alpha \beta} = \frac{1}{3} \phi^2,
\label{WeylForm}
\end{align}
where only the scalar
\begin{align}
\phi
&=\bar{R}+ \boxII\log(S)
\label{PhiEq1}
\\
&=\bar{R}+ \frac{2\boxII\sqrt{S}}{\sqrt{S}} - \frac{2(\nablaII \sqrt{S})^2 }{S} 
\\
&=\bar{R} + 2 \nablaII q
\label{PhiFromQ}
\end{align}
depends on derivatives of the metric \cite{Nozawa:2008rjk}. It is clear from \eqref{PhiFromQ} and \eqref{RfromJ2} that $\phi$ is a total derivative. By contrast, the "reduced Weyl tensor" $\tilde{C}_{\alpha\beta\gamma\delta}$ only depends on metric components directly, and consequently takes on a relatively simple form in any coordinate system (see Appendix D of \cite{Nozawa:2008rjk}). This also means that 
\begin{align}
C_{\alpha\beta\gamma\delta}=0\Leftrightarrow \phi=0.
\end{align}
\ADS{
It is trivial to see that $\phi=0$, hence \ADSxS is conformally flat. 
}

In the remainder of this section, we will prove the statement of box \ref{B0} that if scale invariance is already present, then $\phi=0$ implies full conformal invariance, and vice versa. However, first we establish an identity which will be very useful later.  Since $J^k q_k =1$, 
we have
\begin{align}
0=    q^i \nablaII_i (J^k q_k) &=
  q^i q^k(\nablaII_i J_k) + q^i J^k (\nablaII_i q_k).
\end{align}
Using the Killing equation \eqref{Fequation}, we can show
\begin{align}
  &  q^i q^k (\nablaII_i J_k) = q^{(i} q^{k)} (\nablaII_{[i} J_{k]}) =0,
    \\
 &   \Rightarrow q^iJ^k (\nablaII_i q_k)  = 0.
\end{align}
Starting from \eqref{tij}, we can we can hence determine 
\begin{align}
    q^i t_{ij} J^j = \frac{\bar{R}}{2} + \nablaII q = \frac{1}{2} \phi.
    \label{qtJ}
\end{align}
This result elegantly connects the Weyl tensor, the NEC, and the base space Killing vector $J^i$.

\subsection{Conditions for full Conformal Invariance}
\label{sec::ConformalConditions}

The isometry group of the bulk spacetime is enhanced from scale without conformal invariance to full conformal invariance if vectors of the type \eqref{Ktypes} satisfy the Killing equation in addition to the $D$-type Killing vector \eqref{D-type}. 
In this section, we will hence assume that the Killing equations \eqref{Sequation} and \eqref{Fequation} for the generator of scale transformations are already satisfied, and investigate what additional constraints we get from the Killing equations for the vectors \eqref{Ktypes}. The only additional constraints turn out to be 
\begin{align}
0=2J_i+S\partial_i\mK.
    \label{Kequation}
\end{align}
as well equation \eqref{mKequationLong} which arose from the Lie-algebra, and can be more succinctly written as 
\begin{align}
0=2\mathcal{K}+\bar{\mL}_J\mK.
\label{mKequation}
\end{align}
Essentially, if and only if a scalar $\mK(w,\theta)$ can be constructed that satisfies both \eqref{Kequation} and \eqref{mKequation} for given $h_{ij}$ and $S$, scale invariance is enhanced to full conformal invariance.

\subsection{Conformal Invariance \texorpdfstring{$\Rightarrow \phi=0$}{implies vanishing phi}}
\label{sec::CIphi0}

Let us now assume that an appropriate $\mK$ satisfying \eqref{Kequation} and \eqref{mKequation} exists, and hence the bulk isometry group corresponds to full conformal invariance. From \eqref{Kequation}, we have
\begin{align}
    J_i = -\frac{1}{2} S \nablaII_i \mathcal{K}.
    \label{mKequation2}
\end{align}
Using \eqref{mKequation}, the norm of the 2d-Killing vector can be written as
\begin{align}
    J^i J_i = -\frac{1}{2} S J^i \nablaII_i \mathcal{K} = -\frac{1}{2} S \bar{\mL}_J{\mK} = \mK S. 
\end{align}
Using \eqref{PhiEq1} and \eqref{RfromJ2}, we can hence write
\begin{align}
    \phi &= -\boxII \log(J^2)+ \boxII\log(S) \notag\\ & = -\boxII \log (\mathcal{K} S) + \boxII\log(S) \notag \\
  &= - \boxII \log \mathcal{K}.
  \label{phiFromK}
\end{align}
On the other hand we can write,
\begin{align}
    J^i J_i &= \left(-\frac{1}{2} S \nablaII^i \mathcal{K}\right)\left(-\frac{1}{2} S \nablaII_i \mathcal{K}\right)
    \\
    &\Rightarrow     \left(-\frac{1}{2} S \nablaII^i \mathcal{K}\right)\left(-\frac{1}{2} S \nablaII_i \mathcal{K}\right) = \mK S 
    \\
    &\Rightarrow S = \frac{4 \mK}{\nablaII^i \mK \nablaII_i \mK}
        \\
    &\Rightarrow     J_i = -\frac{2 \mK}{\nablaII^i \mK \nablaII_i \mK} \nablaII_i \mathcal{K}.
    \label{SfrommK}
\end{align}
Let us manipulate \eqref{mKequation2} further to obtain the divergence of the Killing vector as,
\begin{align}
    \nablaII^i J_i &= - \nablaII^i \left(\frac{2 \mK}{\nablaII^j \mK \nablaII_j \mK} \nablaII_i \mK\right) \notag\\ &=
    -2 - \frac{2 \mK \boxII \mK}{\nablaII^i \mK \nablaII_i \mK} + \frac{2 \mK \nablaII^i \mK}{(\nablaII^j \mK \nablaII_j \mK)^2} \nablaII_i (\nablaII^k \mK \nablaII_k \mK) \notag \\&= -2 - \frac{2 \mK \boxII \mK}{\nablaII^i \mK \nablaII_i \mK} - \frac{1}{\nablaII^j \mK \nablaII_j \mK} J^i\nablaII_i (\nablaII^k \mK \nablaII_k \mK).
    \label{nablaJintermediate}
\end{align}
The last term can be simplified considerably via a series of manipulations. We know that the Lie derivative satisfies the following product rule,
\begin{align}
 u^\lambda \nabla_{\lambda}(p^\mu q_\nu)=   \mL_u (p^\mu q_\nu) = \mL_u (p^\mu) q_\nu + p^\mu \mL_u (q_\nu).
\end{align}
Employing this in our calculations, and using that $\bar{\mL}_Jh_{ij}=0$ by assumption \eqref{Fequation}, we obtain
\begin{align}
    J^i \nablaII_i (\nablaII^j \mK \nablaII_j \mK) = \bar{\mL}_J{(\nablaII^j \mK \nablaII_j \mK)} = 2 \nablaII^j \mK \bar{\mL}_J{(\nablaII_j \mK)} .
\end{align}
Since Lie and covariant derivatives commute trivially when acting on a scalar \cite{Tanski2020}, the previous expression simplifies as
\begin{align}
     J^i \nablaII_i (\nablaII^j \mK \nablaII_j \mK) 
    =  2 \nablaII^j \mK \nablaII_j(\bar{\mL}_J{\mK)} 
    &= -4 \nablaII^j \mK \nablaII_j \mK ,
\end{align}
where \eqref{mKequation} has been used. 
Therefore, starting from \eqref{nablaJintermediate} again, we find
\begin{align}
   \nablaII^i J_i &= 2 - \frac{2 \mK \boxII \mK}{\nablaII^i \mK \nablaII_i \mK} 
   \\
  &  = -\frac{2 \mK^2 \boxII \log \mK}{\nablaII^i \mK \nablaII_i \mK} 
  \\& = -\frac{J^i J_i}{2} \boxII \log \mK. 
\end{align}
The LHS of the above equation has to vanish ($\nablaII J =0$) for $J^i$ to be a Killing vector field. On the RHS, since $J^i J_i \neq 0$, we thus find $- \boxII \log \mK = \phi= 0$ as a consequence of \eqref{phiFromK}. This shows that full conformal invariance necessitates the Weyl tensor to vanish.  We will prove the converse statement in the next subsection. 
\\

However, before we do so we briefly mention an alternative but slightly less elegant proof strategy that arrives at the same result, and which was worked out in detail in \cite{Rykala}. The idea is based on the fact that if $\mV^\mu$ is a Killing vector field of the 4d-spacetime, its Lie-derivative will yield zero when acting on \textit{any} curvature tensor defined from the metric, not just the metric itself. This can especially be applied to the Weyl tensor \eqref{WeylAndReducedWeyl}:
\begin{align}
&0=\mL_\mV g_{\mu\nu}
\\
\Rightarrow\ &0=\mL_\mV C_{\mu\nu\alpha\beta}
= (\mL_\mV\phi) \tilde{C}_{\mu\nu\alpha\beta}
+\phi\mL_\mV\tilde{C}_{\mu\nu\alpha\beta}.
\label{PawelMethod}
\end{align}
As we discussed in section \ref{sec::Weyl_form}, in any coordinate system the components of $\tilde{C}_{\mu\nu\alpha\beta}$ tend to be relatively simple, as only $\phi$ depends on derivatives of metric components and in essence absorbs most of the complexity. By comparing the various components for concrete choices of $\mu,\nu,\alpha,\beta$ in equation \eqref{PawelMethod}, we can eliminate the expressions $\mL_\mV\phi$ and $\phi$ from the system of equations (assuming $\phi\neq 0$) and arrive at some relatively simple conditions. These conditions are then necessary, but not sufficient conditions for $\mV$ to be a Killing vector. As it turns out however, these conditions are mutually exclusive for the generators of special conformal transformations \eqref{Ktypes}, meaning that these vectors can only be Killing vectors if \eqref{PawelMethod} is satisfied in the trivial way $\phi=0$.

\subsection{\texorpdfstring{$\phi=0 \Rightarrow $}{Vanishing phi implies} Conformal Invariance }
\label{sec::phi0impliesconformal}

Let us now assume that $\phi=0$ in this subsection and derive what follows from this. With an ansatz for $q^i$ of the form \eqref{qQansatz}, we find  
\begin{align}
  0\equiv  \phi &= \bar{R} + 2 \nablaII_i \left(\frac{J^i}{J^2}\right) + 2 \boxII \log{Q} \\&= -\boxII \log{J^2} + 2 \boxII \log{Q}
      \label{phiFromJ2andQ}
  \\
  &\Leftrightarrow
\boxII \log{\frac{Q^2}{J^i J_i}} =0.
\label{phi=0}
\end{align}
Following the same arguments as in the discussion at the end of section \ref{sec::GBandCurvature}, we can conclude that \eqref{phi=0} implies 
\begin{align}
    2\log{Q} &= const. + \log{J^2} 
    \Rightarrow q_i = \frac{J_i}{J^2} + \frac{1}{2} \nablaII_i (\log{J^2})\,.
    \label{q_for_phi0}
\end{align}
Rearranging this explicit form of $q^i\equiv\frac{\nablaII_i S}{2S} $, we can show
\begin{align}
  2J_i &=-S \nablaII_i\left(\frac{J^2}{S}\right). 
    \label{qforPhi0}
\end{align}
Comparing \eqref{qforPhi0} to \eqref{Kequation}, we see that \eqref{Kequation} is satisfied if 
\begin{align}
\mK = \frac{J^2}{S}.
\label{mKcandidate}
\end{align}
The same candidate function for $\mK$ satisfies \eqref{mKequation} as can be easily verified using \eqref{Sequation} and \eqref{LieJ2}. To summarise, starting from the assumption that $\phi=0$, we have derived a function $\mK$ \eqref{mKcandidate} that satisfies both \eqref{Kequation} and \eqref{mKequation} in this case, meaning that the vectors generating special conformal transformations \eqref{Ktypes} are indeed Killing vectors. Thus, vanishing of the Weyl tensor implies full conformal invariance, completing the proof of the result in box \ref{B0}.

\ADS{
Equation \eqref{mKcandidate} simplifies to $\mK=e^{-\frac{2w}{L}}L^2$. As $\phi=0$, it is not surprising to find that the vectors \eqref{Ktypes} with this choice for $\mK$ and $J$ in \eqref{JforAdSS1} 
are indeed Killing vectors. Consequently, the isometry group of \ADSxS corresponds to the direct product of conformal invariance and the $U(1)$ symmetry of the compact extra dimension.
}

\subsection{\texorpdfstring{$\phi=0 \Rightarrow \det[t_{ij}]=0$}{Vanishing phi implies NEC} }
\label{sec::det_phi}

The vanishing of $\phi$ has a significant additional consequence, as we are about to discuss.  
As seen in the previous subsection, $\phi=0$ implies $Q^2=const.\times J^2$, bringing $q_i$ to the form \eqref{q_for_phi0}. This means that for any base space metric $h_{ij}$ and Killing vector $J^i$, there is a choice of $S$ (and hence $Q$) satisfying \eqref{Sequation} that leads to $\phi=0$. 
Plugging this explicit form \eqref{q_for_phi0} of $q_i$ into \eqref{DETtijq}, a straightforward calculation shows
\begin{align}
    \phi=0 \Rightarrow \det[t_{ij}]=0. 
\end{align}
If in addition $\text{tr}[t_{ij}]\geq 0$ holds, this means that $\phi=0$ (and hence full conformal invariance) implies the NEC to be satisfied.

\subsection{Petrov Classification}
\label{sec::Petrov}

In this section, we discuss the algebraic classification of the Weyl tensor using the Newman–Penrose (NP) formalism \cite{Newman:1962}. This involves projecting the Weyl tensor onto a complex null tetrad \{$l_\mu, k_\mu, m_\mu, \bar{m}_\mu$\}— a complete basis consisting of two real $\{l_\mu, k_\mu\}$ and a complex-conjugate pair $\{m_\mu, \bar{m}_\mu\}$ of null vectors, which satisfy the following properties:
\begin{align}
    &l_{\mu} l^{\mu} = k_{\nu} k^{\nu} = m_{\rho} m^{\rho} = \bar{m}_{\sigma} \bar{m}^{\sigma} = 0\,, \notag \\&
    l_{\mu} m^{\mu} = l_{\nu} \bar{m}^{\nu} = k_{\rho} m^{\rho} = k_{\sigma} \bar{m}^{\sigma} = 0\,, \notag\\&
    -l_{\mu} k^{\mu} =m^{\nu} \bar{m}_{\nu} = 1\,, \notag\\&
    g_{\mu \nu} = -l_{\mu} k_{\nu}-k_{\mu} l_{\nu}+m_{\mu} \bar{m}_{\nu}+\bar{m}_{\mu} m_{\nu}\,, \notag\\&
    g^{\mu \nu} = -l^{\mu} k^{\nu}-k^{\mu} l^{\nu}+m^{\mu} \bar{m}^{\nu}+\bar{m}^{\mu} m^{\nu}\,.
    \label{tetradproperties}
\end{align}
Projecting the Weyl tensor onto this basis yields five complex scalars $\{\Psi_{0}, \Psi_{1}, \Psi_{2}, \Psi_{3}, \Psi_{4}\}$, which collectively encode its independent components. These are known as Weyl scalars and are defined as follows \cite{Newman:1962, Stephani:2003tm} (see also appendix E of \cite{Frolov:1998wf}):
\begin{align}
    &\Psi_{0} \equiv C_{\mu\nu\rho\sigma} l^{\mu} m^{\nu} l^{\rho} m^{\sigma}\,, \;\; \Psi_{1} \equiv C_{\mu\nu\rho\sigma} l^{\mu} k^{\nu} l^{\rho} m^{\sigma}\,, \notag \\&\Psi_{2} \equiv \frac{1}{2} C_{\mu\nu\rho\sigma} l^{\mu} k^{\nu} (l^{\rho} k^{\sigma}-m^{\rho} \bar{m}^{\sigma}),\notag\\&\Psi_{3} \equiv C_{\mu\nu\rho\sigma} l^{\mu} k^{\nu} k^{\rho} \bar{m}^{\sigma}\,, \;\; \Psi_{4} \equiv C_{\mu\nu\rho\sigma} \bar{m}^{\mu} k^{\nu} \bar{m}^{\rho} k^{\sigma}\,.
    \label{generalcomplexscalars}
\end{align}
The classification of the Weyl tensor is based on which of these Weyl scalars vanish and is referred to as the Petrov classification. This serves as a practical method for determining the Petrov type of a spacetime. Another approach, which more closely resembles Petrov’s original method, involves solving an eigenvalue problem after projecting the Weyl tensor onto the space of bivectors (see \cite{Stephani:2003tm} for more details). 

To determine the Petrov type of the four-dimensional bulk spacetime \eqref{Metric_ansatz} under consideration, we construct a complex null tetrad that satisfies \eqref{tetradproperties}.
But before doing so, to simplify the calculations, we can use the fact that every 2D metric is conformally flat, and choose a coordinate system $\{u, v\}$ (without loss of generality) for our base space such that the bulk metric locally takes the following simpler form
\footnote{These are called \textit{isothermal coordinates} on the base space, and can locally be constructed by solving the Beltrami equation. However, it is unclear to us whether in general this construction extends to the entire base space in a single valued manner due to the non-trivial topology. As the Petrov classification is only concerned with local curvature, we do not have to worry about this issue right now. The metric ansatz \eqref{Metric_ansatz} and some of its properties were studied in this coordinate system in \cite{Rykala}.   }:
\begin{align}
    ds^2 = S(u, v)(-dt^2+dx^2)+ f(u, v)(du^2+dv^2).
    \label{metricPetrov}
\end{align}
It is straightforward to check that the following vectors form the desired tetrad for our metric,
\begin{align}
l_\mu=\left(\begin{array}{c}
1\\
1 \\
0 \\
0
\end{array}\right)\,,\ \ 
k_\mu=\frac{1}{2}\left(\begin{array}{c}
S(u, v) \\
- S(u, v)\\
0 \\
0
\end{array}\right)\,, \ \
m_\mu=\frac{1}{\sqrt{2}} \left(\begin{array}{c}
0\\
0\\
-i \sqrt{f(u, v)} \\
\sqrt{f(u, v)}
\end{array}\right)\,, \ \
\bar{m}_\mu=\frac{1}{\sqrt{2}}\left(\begin{array}{c}
0\\
0\\
i\sqrt{f(u, v)} \\
 \sqrt{f(u, v)}
\end{array}\right)\,.
\label{tetrad}
\end{align}
These vectors are not unique and allow rescalings by appropriate scalars as long as they satisfy \eqref{tetradproperties}. Using the Weyl tensor for our metric in the form \eqref{metricPetrov} and the tetrad \eqref{tetrad}, we obtain
\begin{align}
    &\Psi_{0} = \Psi_{1} = \Psi_{3} = \Psi_{4} = 0, \notag\\& \Psi_{2} = - \frac{\phi}{12}.
\end{align}
Thus, our metric ansatz is of Petrov type D for $\phi\neq0$ and type O for $\phi=0$. 
According to the standard physical interpretation due to Szekeres (see \cite{Stephani:2003tm} for details), the Weyl scalar $\Psi_2$ encodes the Coulomb-like component of the gravitational field associated with an isolated source, whereas $\Psi_0$ and $\Psi_4$ describe incoming and outgoing transverse gravitational radiation, respectively, and $\Psi_1$ and $\Psi_3$ correspond to longitudinal radiative modes. Therefore, four-dimensional metrics belonging to the class of \eqref{Metric_ansatz} for which the Weyl tensor is non-vanishing may describe the gravitational field generated by an isolated source. Other important examples of Petrov type D spacetimes include the Schwarzschild, Kerr, and Reissner–Nordström solutions. Deviations from this physical interpretation can arise in spacetimes with cylindrical isometries, which are beyond the scope of the present work (see \cite{Hofmann:2013zea} for more details).

\section{No physical models of Scale without Conformal Invariance}
\label{sec::proofs}

In this section, we will provide a proof for our main conjecture (box \ref{B1}) in subsection \ref{sec::full_proof}. However, before doing so, we will first introduce an additional tool into our toolbox in subsection \ref{sec::beta_scaling} by discussing how rescaling the Killing vector $J$ relates different geometries to each other. Then, in subsection \ref{sec::special_cases}, we will explore various special cases where simplifying assumptions hold, and where our main conjecture can be given alternative proofs. The reason why we introduce these special-case proofs before the general proof in section \ref{sec::full_proof} is threefold:
\begin{itemize}
    \item Firstly, these special-case proofs may be easier to follow and a bit more elegant than the fully general proof. 
    \item Secondly, the special cases in question are very natural, as the comparison to the \ADSxS case shows. 
    \item Thirdly, enumerating these alternative proofs and their underlying methods may be useful for future attempts to find proofs (or construct counterexamples) for similar setups in higher dimensions, or with relaxed topological or energy conditions (see the discussions in section \ref{sec::outlook}).
\end{itemize}

\subsection{Scaling of the Killing Vector}
\label{sec::beta_scaling}

As stated in box \ref{B3}, our construction needs to use a metric on the base space that allows for at least one Killing vector field. Once a base space metric $h_{ij}$ and a Killing vector $J^i$ are picked, a function $S$ solving \eqref{Sequation} has to be found to yield a complete model of the form \eqref{Metric_ansatz}. But if $J$ is a Killing vector, then of course so is $\alpha J$ for some non-zero number $\alpha$. How does this freedom to rescale the Killing vector affect the possible solutions for $S$?
If for a given pair $S$ and $J^i$ equation \eqref{Sequation} holds, then we can also show that
\begin{align}
    -2S^\beta+\bar{\mL}_{\alpha J} S^\beta=0\ \ \text{  if  }\ \     \alpha=\frac{1}{\beta}.
    \label{Jscaling}
\end{align}
This means that rescaling $J$ corresponds to exponentiating $S$, which we can use to make derivative terms of $S$ in \eqref{DETtij} more dominant compared to $\bar{R}$ or less. 
Specifically, under this kind of re-scaling, we find:
\begin{align}
\bar{R}\rightarrow &\bar{R},
\\
q^i\rightarrow &\beta q^i,
\\
\phi
\rightarrow &\bar{R}+ \beta (2 \nablaII q),
\\
    t_{ij} \rightarrow &\frac{\bar{R}}{2} h_{ij} + \beta \left[(\nablaII q) h_{ij} - 2 \nablaII_i q_j \right] + \beta^2 \left[2 q^2 h_{ij} - 2 q_i q_j \right],
  \label{tij_with_beta_and_q}
  \end{align}
  and thus
  \begin{align}
      \Tr[t_{ij}] \rightarrow &\bar{R} +2\beta^2 q^2,
      \\
      \det[t_{ij}] \rightarrow &  \frac{1}{4}\bar{R}^2
+    \beta^2 [\bar{R} q^2 +(\nablaII q)^2 - 2 \nablaII_i q_j \nablaII^i q^j]
+\beta^3\left[ 2 q^4 \nablaII_i \left(\frac{q^i}{q^2}\right) 
\right].
\label{det_with_beta_and_q}
\end{align}
Another useful result will be
\begin{align}
    q^i t_{ij} q^j &\rightarrow \frac{\bar{R}}{2} q^i q_i + \beta (q^i q_{i})^2 \left[\nablaII_k \left(\frac{q^k}{q^j q_j}\right)\right].
    \label{qtq}
\end{align}
Introducing the factor $\beta$ hence allows to order and collect the terms appearing for example in $q^it_{ij}q^j$ or in $\det[t_{ij}]$ in some very insightful ways. As we are about to see in the next subsections, this will facilitate the formulation of simple proofs of our conjecture in certain special cases. However, before we do so we should quickly address the physical meaning or interpretation of this newly introduced factor $\beta$. As we discussed in section \ref{sec::AnsatzAndNotation}, we are interested in metrics of the form \eqref{Metric_ansatz}, which are fully specified by a choice of $h_{ij}$ and $S$. If instead of $S$, we pick $\tilde{S}=S^\beta$ for some $\beta$ while keeping the same $h_{ij}$, this then corresponds to a different spacetime metric. The parameter $\beta$ should hence be understood as an auxiliary parameter, that allows us to study entire families of distinct, but geometrically related metrics at one glance, or to generate potentially more interesting metrics from a given first guess. In the end, however, we should always be able to set $\beta=1$ again by explicitly adjusting the function $S$ as needed.  
\ADS{
For the \ADSxS example, we see from \eqref{JforAdSS1} that the AdS-scale $L$ acts like the factor $1/\beta$ in \eqref{Jscaling}. This explains the appearance of the factor $1/L$ in the exponent of $S=e^{\frac{2w}{L}}$ in \eqref{AdSS1R}. Furthermore, the Ricci scalar $R$ of the 4d bulk spacetime is given in warped product language by the expression
\begin{align}
    R=\bar{R}-4\beta\nablaII q -6\beta^2 q^2=-\frac{6\beta^2}{L^2}.
    \label{R4d}
\end{align}
For the last simplification, we have used \eqref{AdSS1R} and \eqref{AdSS1q}. The large $L$ ($\sim$ small $\beta$) limit which is so important in AdS/CFT hence takes the bulk curvature towards zero only because $\bar{R}=0$. 
}

\subsection{Special cases with no-go theorems}
\label{sec::special_cases}

We will now explore several special cases in which our main conjecture in box \ref{B1} can be proven straightforwardly.



\stoptoc
\subsubsection{Case of \texorpdfstring{$q^2=const.$}{constant norm of q}}

If $q^i q_i$ is a constant, then \eqref{qtq} simplifies to 
\begin{align}
    q^i t_{ij} q^j &= \frac{\bar{R}}{2} q^i q_i + \beta (q^i q_{i}) \nablaII_k q^k=\frac{\phi}{2}q^2.
    \label{qtqconst}
\end{align}
But $q^2>0$ everywhere and $\phi$ is clearly a total derivative (see the discussion in section \ref{sec::GBandCurvature}), and so this is either zero everywhere or negative somewhere. In the latter case, the NEC \eqref{NEC2d} is violated (at least by the vector $q^i$) wherever $\phi<0$ while in the case where $\phi=0$ everywhere, we are dealing with full conformal invariance as explained in section \ref{sec::phi0impliesconformal}.

\subsubsection{Case of \texorpdfstring{$\bar{R}=0$}{vanishing R}}
\label{sec::R0}

Consider now the case where $\bar{R}=0$ everywhere. Then \eqref{qtq} gives us 
\begin{align}
    q^i t_{ij} q^j &=\beta (q^i q_{i})^2 \left[\nablaII_k \left(\frac{q^k}{q^j q_j}\right)\right].
\end{align}
Again, this is a function which is positive everywhere multiplied with a total derivative, hence this expression will have to be negative somewhere (indicating the NEC is violated by the vector $q^i$), or it will be zero everywhere. 
\\

The latter case should be discussed with some care. If both $\bar{R}=0$ and $\nablaII_k \left(\frac{q^k}{q^j q_j}\right)=0$ everywhere, the determinant of $t_{ij}$ simplifies to 
 \begin{align}
      \det[t_{ij}]=  \beta^2 [(\nablaII q)^2 - 2 \nablaII_i q_j \nablaII^i q^j].
      \label{qtq=0}
\end{align}
It is easy to show that this term is non-positive:\footnote{We can always locally choose a coordinate system where $h_{ij}=\delta_{ij}$. Then, for any symmetric tensor $A_{ij}$, we find $A_{ij}A^{ij}=A_{11}^2+A_{22}^2+2A_{12}^2\geq0$.}
\begin{align}
   & \left(\nablaII_i q_j -\frac{1}{2}\nablaII q h_{ij}\right)\left(\nablaII^i q^j -\frac{1}{2}\nablaII q h^{ij}\right)\geq 0,
    \label{ineq1}
    \\
   & \Rightarrow   (\nablaII q)^2  - 2 \nablaII_i q_j\nablaII^i q^j \leq 0.
   \label{2ndorderterm_ineq}
\end{align}
Hence, the only way to preserve the NEC in this special case is for the determinant to vanish everywhere. 
Thus, let us assume that $t_{ij}$ is positive semi-definite with one zero eigenvalue corresponding to the eigenvector $q^i$ (because we need \eqref{qtq=0} to vanish), and one further positive eigenvalue. We can hence assume
\begin{align}
t_{ij} q^j =0,
\end{align}
and with \eqref{qtJ}, this in turn leads to
\begin{align}
0=J^i t_{ij} q^j =\frac{1}{2}\phi.
\end{align}
Once more, we find that for the special case we are investigating, the only situation that avoids a violation of the NEC is the one where $\phi=0$, and hence SwCI is extended to full conformal invariance.

\subsubsection{Case of \texorpdfstring{$q$ and $J$}{q and J} being co-linear}
\label{sec::q||J}

Equation \eqref{qtJ} allows us to identify another special case in which the main conjecture of box \ref{B1} is obviously true: If $q^i=J^i/J^2$ (i.e.~$Q=const.$ in \eqref{qQansatz}), then equation \eqref{qtJ} implies\footnote{Actually,  $q^i=J^i/J^2$ implies $\nablaII q=0$ (see \eqref{DivqQ}), so in this case we can even write $\phi=\bar{R}$.} 
\begin{align}
    q^i t_{ij} q^j=\frac{1}{2J^2} \phi.
\end{align}
As $(q^i t_{ij} q^j) J^2$ is manifestly a total derivative, the same argument as above applies: Either the NEC is violated somewhere, or we are dealing with full conformal invariance.

\ADS{
For AdS$_3\times$S$^1$, we have the special case where  $\bar{R}=0$ and $q^i=J^i/J^2,\ q^2=1/L^2=const.$  This spacetime hence combines all of the special cases discussed in this subsection so far. However, as $\phi=0$, the NEC is saturated and the metric allows for full conformal invariance.  
}

\subsubsection{Limits of large and small \texorpdfstring{$\beta$}{beta}}

In the limit $\beta\rightarrow0$, we clearly see $   t_{ij}\rightarrow\frac{\bar{R}}{2} h_{ij} $ in \eqref{tij_with_beta_and_q}. As $\bar{R}$ is a total derivative, it is either zero everywhere or negative somewhere. In the latter case, the NEC is clearly violated as $t_{ij}$ is negative definite wherever $\bar{R}<0$. In the case where $\bar{R}=0$ everywhere, the NEC is saturated, but also $\phi=\bar{R}=0$ and so SwCI is enhanced to full conformal invariance. 
\\

The limit $\beta\rightarrow\pm\infty$ is very similar to the case $\bar{R}=0$, as for large enough $|\beta|$ the determinant \eqref{det_with_beta_and_q} is dominated by the leading order term $2 q^4 \nablaII_i \left(\frac{q^i}{q^2}\right) $ which, containing a total derivative, will either be negative somewhere or zero everywhere. If this term vanishes, the order $\beta^2$ term in \eqref{det_with_beta_and_q} becomes the leading order term, and by \eqref{2ndorderterm_ineq} this term will also be negative somewhere unless $\bar{R}=0$ everywhere which is a special case we have already checked. 
\\

This shows that generically, if counter-examples to our main conjecture existed, they would have to fall into some kind of Goldilocks-zone of intermediate values of not too small and not too big $\beta$ (for a given base space metric and $S$). As explained above, $\beta$ can be understood as an auxiliary parameter that allows us to study entire families of distinct, but geometrically related metrics at one glance. We will quickly illustrate this on some simple examples. Consider a model where
\begin{align}
&h_{ww}=f_0(\theta)\equiv 1+\frac{1}{2}\sin (\theta), h_{w\theta}=0, \text{ and } h_{\theta\theta}=1/f_0(\theta)
\Rightarrow h=1,
\\
& \sqrt{S}=e^w f_S(\theta). 
\end{align}

\begin{figure}
    \centering
    \includegraphics[width=0.485\linewidth]{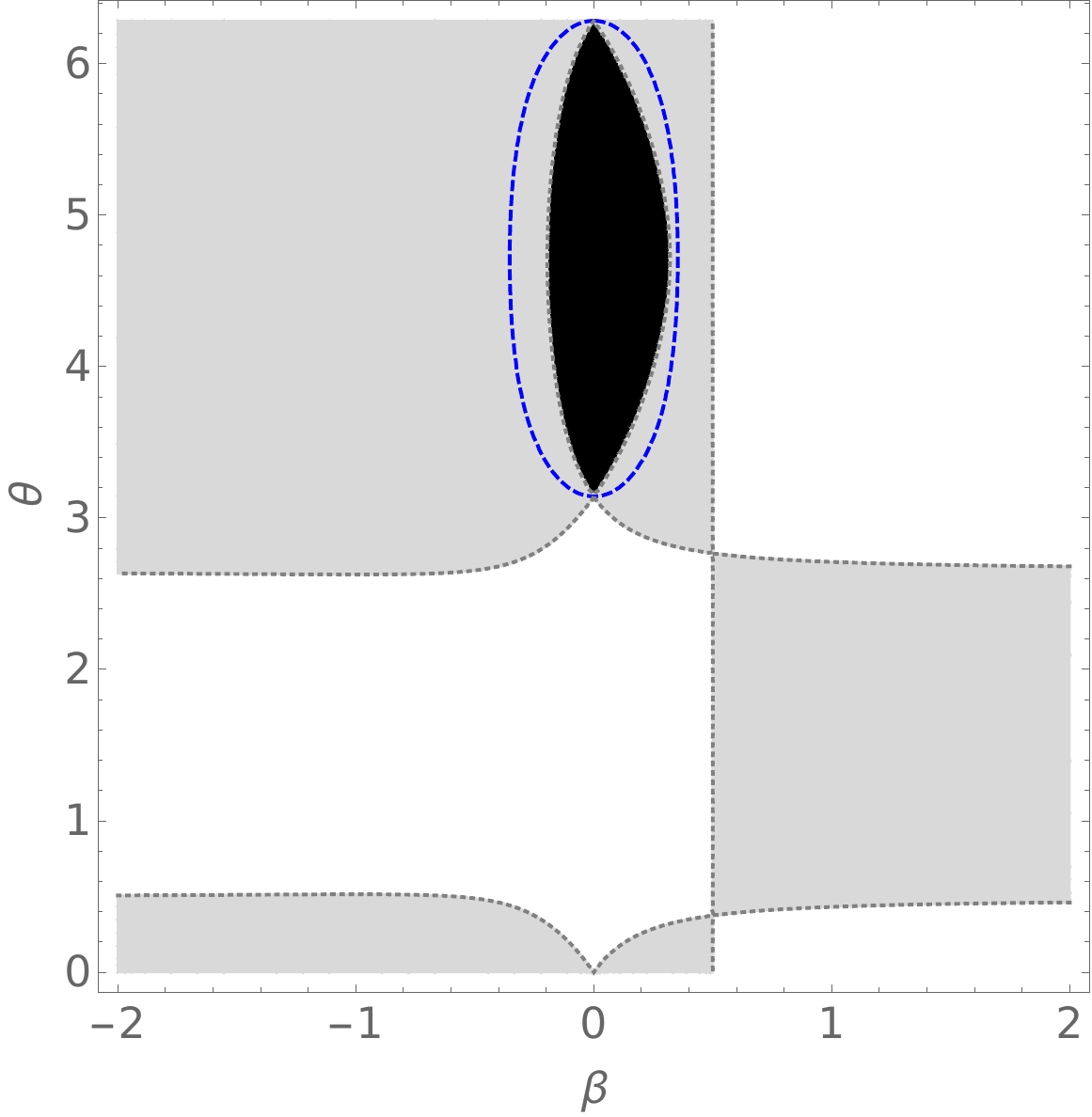}
    \includegraphics[width=0.485\linewidth]{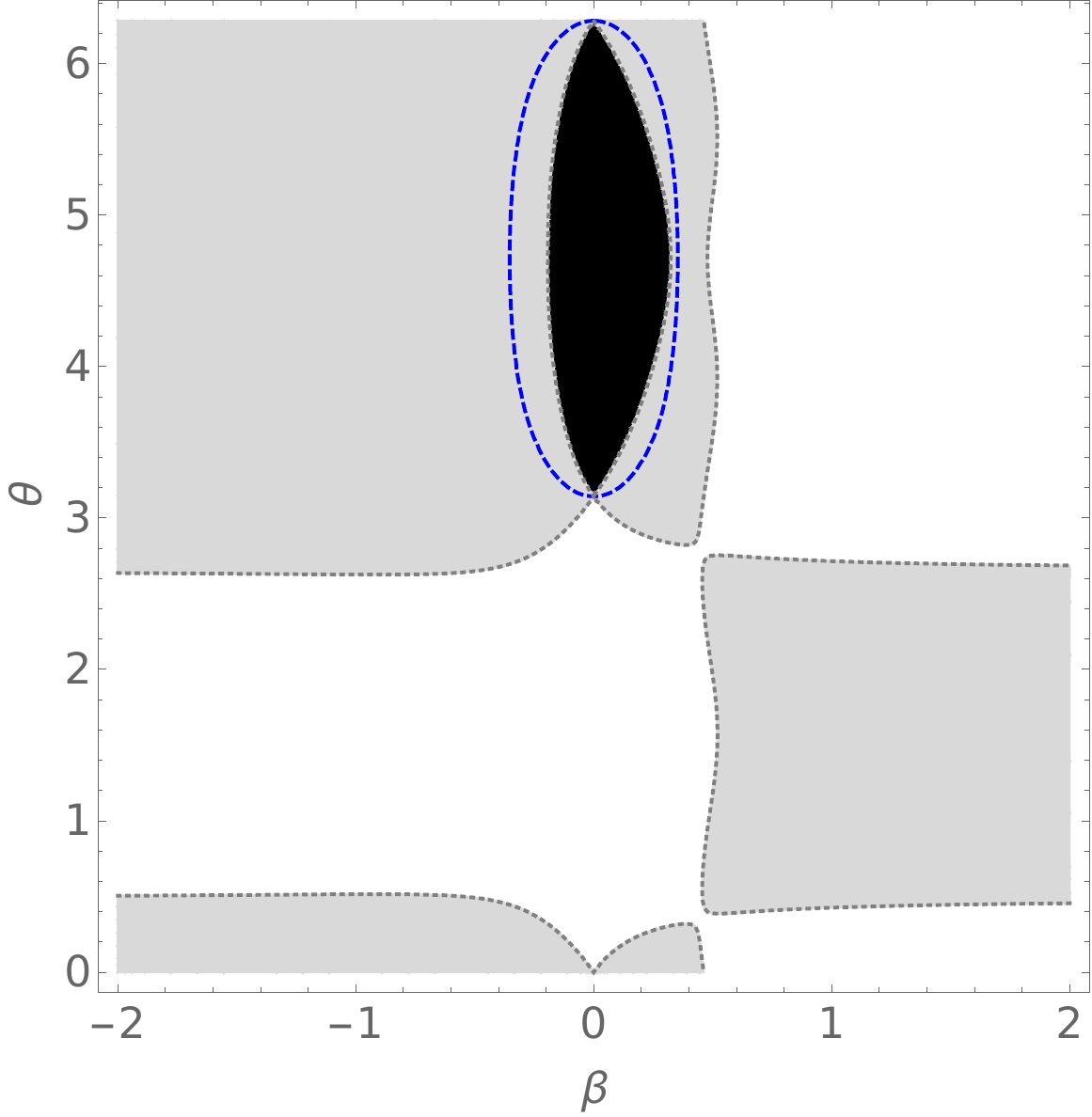}
    \caption{Phase diagrams in the $\beta-\theta-$plane for $f_S(\theta)=f_0(\theta)$ (left) and $f_S(\theta)=f_0(\theta)-\frac{1}{400} \cos ^2(2 \theta)$ (right). These diagrams show domains where $t_{ij}$ is negative-definite (black), indefinite (grey), or positive definite (white). The boundaries of the domains are highlited by dotted grey lines and correspond to the zeroes of $\det[t_{ij}]$ \eqref{det_with_beta_and_q} which is a third order polynomial in $\beta$. The dashed blue line signifies the location where $\Tr[t_{ij}]$ \eqref{tij_with_beta_and_q} vanishes. Many of the lessons of this subsection can be understood quite intuitively from such phase diagrams. For instance, wherever $\bar{R}<0$, a negative-definite domain (black) will appear attached to the $\theta$-axis, contained within a roughly oval shaped contour where $\Tr[t_{ij}]=0$. Hence the NEC will obviously be violated for values of $|\beta|$ too close to zero unless $\bar{R}=0$ everywhere. We also see in the left picture that $\det{[t_{ij}]}=0$ at $\beta=1/2$ where $\phi=0$, as discussed in section \ref{sec::det_phi}.  }
    \label{fig::phase_diag}
\end{figure}

In figure \ref{fig::phase_diag}, we show "phase diagrams" for two examples of $f_S(\theta)$. By "phase diagrams", we mean plots that signify whether the matrix $t_{ij}$ \eqref{tij_with_beta_and_q} is positive-definite, negative-definite, or indefinite at any given point in the $\beta-\theta-$plane. A counter-example to our conjecture in box \ref{B1} would correspond to such a diagram where there exists a vertical section at any value of $\beta$ that does not cross into a domain labelled as indefinite or negative-definite (touching the boundary of a domain is allowed as long as $t_{ij}$ is positive-semi-definite there). 
In the first example, $f_S(\theta)=f_0(\theta)$ and $\phi=0\Rightarrow\det[t_{ij}]=0$ (see section \ref{sec::det_phi}) at $\beta=\frac{1}{2}$, i.e.~there is a situation where the NEC is satisfied, but the corresponding geometry also exhibits full conformal invariance. No other value of $\beta$ avoids a violation of the NEC. The second example $f_S(\theta)=f_0(\theta)-\frac{1}{400} \cos ^2(2 \theta)$ is a small perturbation of the first one. It can be seen from the figure with one glance that there is no value of $\beta$ for which a violation of the NEC is avoided. However, this example is noteworthy as for some finite values of $\beta$ around $\approx\frac{1}{2}$, $q^i t_{ij} q^j$ is indeed positive for all $\theta$. This means that, unlike in the other special cases above, $q^i$ is not always the vector that violates the NEC. Finding a vector for which the NEC is always violated somewhere will be the key to the full proof presented in the next subsection.

\resumetoc
\subsection{General proof}
\label{sec::full_proof}

As established in section \ref{sec::det_phi}, vanishing of $\phi$ implies vanishing of $\det[t_{ij}]$. We might hence try to construct a counter-example to our main conjecture by slightly perturbing a background where $\phi=0$. Firstly, setting $\phi=0$ necessitates
\begin{align}
 q^i &= \frac{J^i}{J^2} + \frac{1}{2 \beta} \nablaII^i \log J^2, \\ 
 q^2 &= \frac{1}{J^2} + \frac{1}{4 \beta^2} \nablaII_i \log J^2 \nablaII^i \log J^2.
\end{align}
where compared to \eqref{q_for_phi0} we have now also introduced the scaling factor $\beta$ explicitly.
Using this, we can check that\footnote{
To show the first line, we use $J^i \nablaII_i (J^j q_j) = 0\Rightarrow J^i J^j(\nablaII_i q_j) = \frac{1}{2} q^j \nablaII_j J^2$ where the Killing equation \eqref{Fequation} has been employed. 
}:
\begin{align}
\phi=0&\Rightarrow  
\\
J^i t_{ij} J^j  &=\left(\frac{\bar{R}}{2} +\beta (\nablaII \cdot q)\right) J^2 - \beta q^j\nablaII_j J^2 + \beta^2 (2 q^2 J^2 - 2) \\ &= \frac{\phi}{2}J^2 - \beta \left(\frac{J^j}{J^2} + \frac{1}{2 \beta} \nablaII^j \log J^2\right) \nablaII_j J^2 + 2 \beta^2 J^2\left(\frac{1}{4 \beta^2} \nablaII_i \log J^2 \nablaII^i \log J^2\right) 
    \\
    &=-\beta\frac{1}{J^2}\mL_{J}J^2=0.
\end{align}
This shows that $J^i$ is the null-eigenvector whenever $\phi=0$ and hence $\det[t_{ij}]=0$\footnote{In section \ref{sec::R0}, we constructed a case where $\det[t_{ij}]=0$ and $q^i$ is the null-eigenvector. This is not a contradiction, as in this special case $q^i$ and $J^i$ were actually co-linear.  }. Now we add a (not necessarily small) perturbation to $q^i$ which would correspond to a change of $S$ while keeping $h_{ij}$ fixed: 
\begin{align}
     q^i &\equiv q^{(0)i}+\gamma^i
     \label{q_ansatz_gamma}
     \\
     &\text{  with  } q^{(0)i} = 
        \frac{J^i}{J^2} + \frac{1}{2 \beta} \nablaII^i \log J^2\ \text{ and 
}J^iq_i=1\Rightarrow \ J^i \gamma_i = 0.
\nonumber
\end{align}
Now, we find
\begin{align}
      J^i t_{ij} J^j &= \beta (J^2)^2 \nablaII_j \left(\frac{\gamma^j}{J^2}\right) + 2 \beta^2 (\gamma^2 + 2 q^{(0)} \cdot \gamma) J^2 
      \\
      &=\beta (J^2 \nablaII_j \gamma^j - \gamma^j \nablaII_j J^2) + 2 \beta^2 \gamma^2 J^2 +2 \beta \gamma^i \nablaII_i J^2 \\
      &= \beta \nablaII_i (\gamma^i J^2) + 2 \beta^2 \gamma^2 J^2. 
      \label{JtJsimplified}
\end{align}
Clearly, the first term includes a total derivative while the second term is manifestly non-negative. For infinitesimal perturbations where $\gamma^2\ll1$, this hence shows that the NEC will be violated unless $\nablaII_i (\gamma^i J^2)=0$ everywhere. 
\\

But as we will now demonstrate, we do not have to assume $\gamma^2\ll1$ at all to show that \eqref{JtJsimplified} will always have to be negative somewhere. 
Following the discussion from the end of section \ref{sec::GBandCurvature}, we adapt a suitable coordinate system, in which 
\begin{align}
    J^i = \left(
\begin{array}{c}
1 \\
 0  \\
\end{array}\right), \quad h_{ij}=
\left(
\begin{array}{cc}
 J^2 (\theta) & h_{w\theta} (\theta) \\
 h_{w\theta} (\theta) & h_{\theta\theta} (\theta) \\
\end{array}\right), \quad h^{ij}= \frac{1}{h}
\left(
\begin{array}{cc}
 h_{\theta\theta} (\theta) & -h_{w\theta} (\theta) \\
 -h_{w\theta} (\theta) & J^2 (\theta) \\
\end{array}\right),
\end{align}
and all functions of $\theta$ have to be periodic due to the cylindrical topology of the base space. 
Comparing \eqref{q_ansatz_gamma} with \eqref{qQansatz}, the vector $\gamma_i$ has the following form:
\begin{align}
\label{vectpertb}
    \gamma_i & 
    \equiv \nablaII_i \log \tilde{Q}, \quad \text{with} \; \tilde{Q} = const. \times\sqrt{\frac{Q^2}{J^{2/\beta}}}  \\ \implies & \gamma_w=0,\ \gamma_\theta = \partial_\theta \log \tilde{Q}\equiv \frac{1}{\beta}W(\theta).
\end{align}
Clearly, $\tilde{Q}$ must be a function of $\theta$ only, because $0=J^i\gamma_i=\bar{\mL}_J \log(\tilde{Q})$. Using the identity
\begin{align}
\nablaII_i v^i=\frac{1}{\sqrt{h}}\partial_i \left(\sqrt{h}v^i\right)    
\end{align}
for any vector $v^i$, \eqref{JtJsimplified} explicitly simplifies to 
\begin{align}
   J^i t_{ij} J^j &=\frac{J^4}{h}\partial_\theta W + \frac{W}{\sqrt{h}}\partial_\theta\frac{J^4}{\sqrt{h}}+\frac{2 J^4}{h}W^2.
   \label{JtJ_W}
\end{align}
As $W$ is the derivative of a periodic function in $\theta$, it is either zero everywhere or negative somewhere. The former case is not interesting, as then $\phi=0$ remains true and full conformal invariance is not broken. In the latter case, we assume $W$ to be a smooth function the graph of which has to cross from positive to negative values somewhere. 
Without loss of generality, we assume this happens at $\theta=0$. The easiest case is the one where $W(0)=0,W'(0)<0$, as it implies
\begin{align}
   J^i t_{ij} J^j\big|_{\theta=0} &=\frac{J^4}{h} W'(0)<0
\end{align}
and hence a violation of the NEC at this exact location. More generally, we can assume
\begin{align}
    W(\theta)=-W_0 \theta^p+\mO(\theta^{p+1}), 
\end{align}
with some $W_0>0$ and an odd positive power $p$. In this case, the leading order behaviour of \eqref{JtJ_W} around $\theta=0$ is given by 
\begin{align}
  J^i t_{ij} J^j&\approx -\left(\frac{J^4}{h}\Big|_{\theta=0}\right)  W_0 p \theta^{p-1} -\frac{W_0}{\sqrt{h}\big|_{\theta=0}}\theta^p\left(\partial_\theta\frac{J^4}{\sqrt{h}}\Big|_{\theta=0}\right)+\frac{2 J^4}{h}\Big|_{\theta=0}W_0^2\theta^{2p}+...
  \\
  & \approx -\left(\frac{J^4}{h}\Big|_{\theta=0}\right)  W_0 p \theta^{p-1}+\mO(\theta^{p}).
\end{align}
Here we have used that $J^4$ and $h$ are positive everywhere and assumed to be smooth functions. Clearly, for some value of $\theta$ sufficiently close to $\theta=0$, the leading order term which is negative dominates and we find $ J^i t_{ij} J^j<0$ even though $ J^i t_{ij} J^j\big|_{\theta=0}=0$. This concludes our proof for all analytic functions $W$. 
\\

What if $W$ is smooth but not analytic, e.g.~if locally around $\theta=0$ it looks like $-\theta e^{-1/\theta^2}$? Let us make the following assumptions that will cover even such cases: $W(0)=0$ and there is some small but finite $\epsilon$ such that 
\begin{align}
&W(-\theta)>0\text{  and  }W(\theta)<0 \text{  for any  }0<\theta<\epsilon,
\\
&W'(\theta)<0 \text{  for any  }0< |\theta|<\epsilon.
\end{align}
Furthermore, we assume
\begin{align}
   & \lim_{\theta\rightarrow 0} \frac{W(\theta)}{W'(\theta)}=\lim_{\theta\rightarrow 0} \frac{1}{\partial_\theta \log(W(\theta))}=0.
\end{align}
The latter assumption is reasonable but important, because it means that $W(\theta)$ goes to zero faster than $W'(\theta)$. Because of this, taking into account that $J^4$ and $h$ are everywhere positive and smooth functions, the leading order contribution as $\theta=0$ is approached will come from the term including $W'$ in \eqref{JtJ_W}: 
\begin{align}
  J^i t_{ij} J^j
  & \approx -\underbrace{\left(\frac{J^4}{h}\Big|_{\theta=0}\right)}_{>0}  W'(\theta) + \text{subleading terms}
  \\
  &<0 \text{  for any  }0< |\theta|<\epsilon.
\end{align}
This extends our proof to the case of smooth but non-analytic functions $W(\theta)$.
\\

To summarise, we have shown that as a consequence of the cylindrical topology of the base space (which necessitates all well-defined quantities on the base space to be periodic in some angular coordinate), the vector\footnote{As a reminder, the existence of this Killing vector $J^i$ is a requirement for the presence of scale invariance (see equation \eqref{Fequation} and box \ref{B3}) and thus the first half of the "local geometrical condition" in box \ref{B1}.
} $J^i$ will always violate the NEC (\eqref{NECforn}, \eqref{NEC2d}) somewhere, unless $\phi=0$ everywhere. But $\phi=0$ everywhere would imply full conformal invariance due to the $\Rightarrow$ direction of the statement in box \ref{B0}, and thus violate the second half of the "local geometrical condition" in box \ref{B1}. This proves the conjecture in box \ref{B1}.

\section{Outlook}
\label{sec::outlook}

In this section, we will discuss several ways in which our present work might be generalized or extended.

\subsection{Relaxation of the topology}
\label{sec::outlook_topology}

One of the main ingredients in our proof was the assumption that the warped extra dimension should be compact, i.e.~that the base space should have the topology of a cylinder. This seems natural, as in top-down holography we often times deal with bulk (product-) spacetimes of a form AdS${}_{d}\times X$ with some compact space $X$. As the scale factor $e^{\frac{2w}{L}}$ (see \eqref{AdSS1q}) diverges as the boundary is approached while the size of $X$ stays the same, the CFT is understood to live on the $d-1$-dimensional boundary of AdS${}_d$, independently of $X$. This is why for an ansatz like \eqref{Metric_ansatz}, we expect the putative dual to live on a Minkowski space $\eta_{ab}$ with coordinates $x^a\in\{t,\vec{x}\}$. There are two ways in which we may relax the condition of the base-space having cylindrical topology:
\\

Firstly, we may allow the base space to have the topology of a (half-)plane, i.e.~$\theta\in\mathbb{R}$. Then, we expect the putative dual theory to live on a space parametrized by coordinates $x^A\in\{t,\vec{x},\theta\}$. For concreteness, let's consider a four-dimensional bulk, which would then be dual to a three dimensional field theory. The vaccum state of this theory would have three-dimensional conformal invariance, where the bulk is AdS${}_4$ and $\theta$ is a spacial direction like any other. We may now imagine a situation where this theory develops a defect localised at $\theta=0$, such that translation invariance in the $\theta$-direction is broken, and the state only retains the symmetries corresponding to SwCI in 2-dimensions (with coordinates $t,x$). The geometry of the bulk spacetime may resemble a so-called Janus solution \cite{Aharony:2003qf,Bak:2003jk,DHoker:2007zhm}. We have not systematically searched for such bulk metrics. See however \cite{Nakayama:2012ed} for results that may be applicable to such defect models. 
\\

Secondly, we may introduce \textit{end-of-the-world (EOTW) branes} in the style of AdS/BCFT models \cite{Takayanagi:2011zk,Fujita:2011fp,Nozaki:2012qd}, such that the allowed range of $\theta$ becomes a finite interval; $\theta_1\leq\theta\leq\theta_2$ without periodic identification. In that case, starting from any model with cylindrical topology, we may use these EOTW-branes to simply cut off the regions where the NEC is violated. E.g~in the models visualized in figure \ref{fig::phase_diag}, we may choose $\beta=1$ and $\theta_1=\pi,\ \theta_2=2\pi$. The problem with this strategy is that the branes have their own world-volume matter content which is usually (but not always \cite{Nakayama:2012ed,Erdmenger:2014xya,Kanda:2023zse}) modelled by a constant tension \cite{Takayanagi:2011zk,Fujita:2011fp,Nozaki:2012qd}. This matter content may then be subject to its own energy conditions, which in turn restrict how these branes can be embedded into the ambient spacetime \cite{Takayanagi:2011zk,Nakayama:2012ed,Erdmenger:2014xya}. We hence conjecture without proof that such EOTW-brane-based bulk models would only be able to avoid NEC-violations in the 4d-bulk spacetime by introducing brane-matter violating the NEC on the brane worldsheet.

\subsection{Synge g-method, spacetime (reverse-)engineering and geometrization}
\label{sec::Synge}

Our strategy in this paper has been to construct a spacetime with certain geometrical properties which we considered to be interesting or desirable (ansatz \eqref{Metric_ansatz} with SwCI isometry group and cylindrical base space), and then check the physicality of this spacetime by computing the energy momentum tensor assuming Einstein's equations and using the NEC as a criterion. This approach to dealing with Einstein's equations has sometimes been referred to as the \textit{Synge g-method} after \cite{synge1960relativity}, 
and is often critically compared with the more traditional approach of making an ansatz for both matter content and spacetime symmetries and then solving Einstein's equations as a set of coupled PDEs. This approach is a common way to consider various science-fiction inspired concepts like wormholes \cite{Visser:1995cc}, warpdrives \cite{Santiago:2021aup} or tractor beams \cite{Santiago:2021xjg} from a hard-science perspective, and thus maybe the term of \textit{spacetime (reverse-)engineering} (following \cite{Santiago:2021xjg}) would be more appropriate. 
\\

Whatever one may call it, the strategy is fundamentally the same in those papers and ours, and so we might ask what we could learn from this body of literature, or vice versa. 
Checking the NEC (or WEC) for an exactly given metric may seem trivial, but for generic ans\"atze, where entire functions of the coordinates appearing in the metric may be left unspecified a priori (like $S(w,\theta)$ was in our example), checking the NEC (or WEC) for the entire class of metrics at once may be non-trivial (see e.g.~\cite{Santiago:2021aup}). In our work, we were able to establish relatively good control over the NEC thanks to the many isometries and the warped product structure \eqref{Metric_ansatz} which allowed us to break everything down to a two dimensional problem and the \textit{relatively} easy to check condition \eqref{NEC2d}. Whether this trick can be applied to other spacetime engineering problems would have to be seen on a case by case basis.
 Considering a transfer of methods in the other direction, there have been some exciting developments concerning numerical tools for the analysis and optimization of warp drive geometries \cite{Helmerich:2024kaf,Helmerich:2024tzr,Fuchs:2024mbm}. If we were to extend our ansatz \eqref{Metric_ansatz} e.g.~to higher base space dimensions or different topology, it might be interesting to see whether such numerical toolkits can be adapted in order to search (by hand or in some automated way) for physical metrics exhibiting SwCI. 
\\

Suppose a physical metric with desirable properties (such as SwCI) were to be found under more general conditions than the ones where our no-go theorem applies. Then, the next step would be to not only show that the NEC is satisfied, but to show what exact matter configuration gives rise to the necessary energy-momentum-tensor. This task is known as \textit{geometrization}, and some broad results can be found in \cite{Krongos:2015fta}. One may also replace the NEC with such geometrisation conditions, and try to prove e.g.~statements on whether or not SwCI can be realised in a spacetime sourced by only a scalar field or only a gauge field for instance. We can already deliver one statement of this type (see appendix \ref{sec::scalar} for a detailed discussion): If the spacetime is sourced by a scalar field, then we will necessarily have $\phi=0$ and thus SwCI will always be extended to full conformal invariance. Another no-go result for when the bulk matter is described by the specific low energy bosonic action of String/M-theory under a certain gauge condition is already given in \cite{Nakayama:2009fe,Nakayama:2010zz}.

\subsection{Relaxation of the NEC}
\label{sec::outlook_ECs}

Of course, we might also consider relaxing the NEC as defining test of the physicality of a spacetime. While the NEC is satisfied by most classical matter configurations, it is well known that it can be violated by quantum effects such as the Casimir effect \cite{Curiel:2014zba,Visser:1995cc}. In realistic systems, the negative Casimir energy tends to be literally outweighed by the positive mass of the plates needed to generate the effect \cite{Cerdonio:2014vua,Visser:1995cc}. However, there may be situations where such quantum effects produce enough NEC-violating stress-energy to support otherwise forbidden spacetime geometries \cite{Maldacena:2018gjk}. Consequently, instead of strictly enforcing the NEC\footnote{Yet another possibility would be to formally keep the NEC, but assume the bulk-gravity theory to be modified, for example by higher curvature terms in the action. See also \cite{Nakayama:2010wx} in this context. }, our strategy may be to use something like the averaged null energy condition (ANEC) \cite{Curiel:2014zba,Visser:1995cc} or the quantum null energy condition (QNEC) \cite{Bousso:2015mna}, or to simply minimize NEC-violation.
In the latter case, we would have to first specify in \textit{which exact way} the NEC violations are to be minimised. We may seek to minimize:
\begin{itemize}
    \item the spacial extend of the region in which the NEC is violated,
    \item the total amount of negative energy present in this region,
    \item the ratio between the amount of negative energy in the region where the NEC is violated and the amount of positive energy in the region where the NEC is satisfied,
    \item the number of directions (or the size of the cone of vectors) in which the NEC is violated. 
\end{itemize}
or any other condition \cite{Moghtaderi:2025cns}. In our setups, such a minimisation may easily be achieved by starting out with a NEC-satisfying metric where $\phi=0$, and just adding a small NEC-violating perturbation to it (see e.g.~figure \ref{fig::phase_diag}). However, in the resulting geometry, the full conformal invariance would only be broken slightly, and it may be desirable to quantify and compare the degrees to which both NEC and full conformal invariance are violated in some way.

\subsection{Matters of dimensionality}
\label{sec::dimensionality}

Starting from section \ref{sec::AnsatzAndNotation}, we considered both the base space and the fiber to have two dimensions for the sake of concreteness. In the appendices \ref{sec::1Dfiber} and \ref{sec::higherDfiber}, we will demonstrate explicitly that our results are actually independent of the number of boundary/fiber-space dimensions except for the special case $n=1$, for which our arguments cannot rule out the existence of bulk models exhibiting scale without conformal invariance. 
As the dimension of the fiber corresponds to the dimension of the putative holographic dual, our main conjecture in box \ref{B1} is thus true for holographic models of field theories in any dimension $n\geq 2$, as long as there is only one warped extra dimension in the bulk.
\\

This continues a trend in holographic studies of RG-flows that has already been remarked on e.g.~in \cite{Nakayama:2009fe}: On the field theory side, many results are particular to a certain dimension and do not generalise easily to other dimensions, such as for example the famous $c$-theorem \cite{Zamolodchikov:1986gt} and the no-go result on SwCI in $n=2$ \cite{Polchinski:1987dy}, as well as many other of the field theory results reviewed in section \ref{sec::Overview}. In contrast, holographic results proven from the bulk perspective, such as the holographic $c$-theorem of \cite{Freedman:1999gp}, often times have a straightforward generalisation to arbitrary dimension of the putative dual. 
\\

Generalising our results to higher dimensions of the base space is a different matter entirely. Throughout our paper we have made use of so called \textit{dimensionally dependent identities} (see also \cite{Lovelock_1970}), i.e.~geometrical statements that are only valid in a specific dimension. This happened for example in our derivation and use of equation \eqref{RfromJ2}, or 
the derivation of the special structure of the Weyl tensor \eqref{WeylForm} \cite{Nozawa:2008rjk}. The latter is at the heart of sections \ref{sec::CIphi0} and \ref{sec::phi0impliesconformal} which together show that for our type of bulk-ansatz, if scale invariance is assumed, then it is extended to full conformal invariance if and only if $\phi=0$. Obtaining this result may have seemed like a somewhat technical detour to some readers, but it was absolutely crucial -- without such a clear and easy to check criterion\footnote{With "clear and easy to check", we especially mean that $\phi=0$ is a simple algebraic condition, while the conditions from which we started in section \ref{sec::ConformalConditions} depend on whether a given set of differential equations allow for a solution with certain properties, and so are comparably indirect.} to distinguish scale from full conformal invariance, any attempt to find a general proof for the main conjecture in box \ref{B1} would likely get stuck in an endless sequence of seeming counter-examples which only turn out to be invalid upon closer inspection on a case by case basis. 
A strategy to find a proof of our general main conjecture\footnote{As discussed in section \ref{sec::Overview}, a similar result exists which seems valid for arbitrary base space dimension, but which is based on strong additional assumptions about the metric and bulk matter content \cite{Nakayama:2009fe,Nakayama:2010zz}.} for a higher number of warped extra dimensions should hence most likely start by trying to generalise the results of sections \ref{sec::CIphi0} and \ref{sec::phi0impliesconformal} first. It is interesting to note that in our proof presented in section \ref{sec::full_proof}, it was the base-space vector $J^i$ that lead to the violation of the NEC. We have tried to keep things as covariant as possible, but qualitatively $J^i$ should be understood as a vector that points "straight" from the boundary into the bulk. In  a more common AdS-like coordinate system, this would correspond to a null-vector in the $t-z-$plane. We find this noteworthy, because in the earlier papers \cite{Nakayama:2009fe,Nakayama:2010zz}  it was also specifically the NEC for a null-vector in the $t-z-$plane that led to the results presented there. This may give an indication for what a good ansatz might be to prove even more general results in the future.




\acknowledgments

We are grateful to Pawe\l\ Ryka\l a for initial collaboration resulting in \cite{Rykala}. For many useful discussions and suggestions, we would like to thank Martin Ammon, Alice Bernamonti, Aranya Bhattacharya, Anatoly Dymarsky, Federico Galli, Falk Hassler, M. Hortaçsu, Cynthia Keeler, Ren\'e Meyer, Christian Northe, Niels Obers, Emanuele Panella, Zahra Raissi, Karl-Henning Rehren, Emiliano Rizza, and Konstantinos Siampos. 
The work of both LC and MF was supported by the Polish National Science Centre (NCN) grant 2021/42/E/ST2/00234. Some of the calculations for this paper were performed using Wolfram Mathematica, and especially the packages diffgeo.m by Matthew Headrick \cite{diffgeo} and xAct \cite{xAct}. The files necessary to reproduce our work are openly available at \cite{UJ/J9M1LS_2025}.


\appendix

\section{Curvature from conformal Killing vectors in 2 dimensions}
\label{sec::conformal_killing_R}

As we know how the Ricci scalar transforms under Weyl transformations, it is possible to write down an equation for the Ricci curvature in terms of the norm of a conformal Killing vector and the appropriate Weyl scale factor which holds for any 2d metric at least locally. 
To this end, suppose we have two metrics $\tilde{h}$ and $h$, related by a Weyl rescaling
\begin{align}
\tilde{h}_{ij}=e^{2\omega}h_{ij}. 
\end{align}
Assuming now that $J$ is a conformal Killing vector for $\tilde{h}$ and that the rescaling $\omega$ is chosen such that it is a Killing vector for $h$ (such choices are always possible), we have 
\begin{align}
\bar{\mL}_J h_{ij}&=0
\\
\bar{\mL}_J \tilde{h}_{ij}& = \lambda \tilde{h}_{ij};\ \ \ \lambda=\tilde{\bar{\nabla}}J.
\end{align}
Combined, this allows us to derive
\begin{align}
\bar{\mL}_J \tilde{h}_{ij}& =\bar{\mL}_Je^{2\omega}h_{ij}=2\tilde{h}_{ij}\bar{\mL}_J\omega
\\
&\Rightarrow \tilde{\bar{\nabla}} J=\lambda = 2 \bar{\mL}_J\omega.
\end{align}
To proceed, we start with the well-known formula for the relation of the curvature scalars in 2d under a Weyl rescaling, and then plug in results like \eqref{RfromJ2} and simplify further:\footnote{Here, subscripts like $\bar{\square}_h$ or $J^2_{\tilde{h}}$ signify whether a given expression is evaluated using the metric $h$ or $\tilde{h}$. }
\begin{align}
\tilde{\bar{R}}&=e^{-2\omega}\left(\bar{R}-2\boxII_h\omega\right)
\\
&=e^{-2\omega}\left(-\boxII_h\log(J^2_h)-2\boxII_h\omega\right)
\\
&=e^{-2\omega}\left(-\boxII_h\log(J^2_{\tilde{h}} e^{-2\omega})-2\boxII_h\omega\right)
\\
&=e^{-2\omega}\left(-\boxII_h\log(J^2_{\tilde{h}})\right)
\\
&=-e^{-2\omega}\left(\frac{1}{\sqrt{h}}\partial_i \sqrt{h}h^{ij}\partial_j \log(J^2_{\tilde{h}})\right)
\\
&=-e^{-2\omega}\left(\frac{e^{2\omega}}{\sqrt{\tilde{h}}}\partial_i \sqrt{\tilde{h}}\tilde{h}^{ij}\partial_j \log(J^2_{\tilde{h}})\right)
\\
&=-\square_{\tilde{h}}\log(J^2_{\tilde{h}}).
\label{RfromconformalJ2}
\end{align}
Essentially, \eqref{RfromJ2} holds exactly, even if $J$ is only a conformal Killing vector.

\section{Weak Energy Condition}
\label{sec::WEC}

Let us also consider the \textit{weak energy condition} (WEC) following basically the same steps as for the NEC in section \ref{sec::NEC}. The WEC reads \cite{Curiel:2014zba}  
\begin{align}
T_{\mu\nu}n^\mu n^\nu\geq0
\end{align}
for \textit{timelike} vectors $n_\mu n^\mu =-1$. 
Splitting $n^\mu$ like in \eqref{null-vector}, we find
\begin{align}
-1=n^\mu g_{\mu\nu} n^\nu = S \hat{n}_a\hat{n}^a+\bar{n}^i h_{ij}\bar{n}^j \Rightarrow \hat{n}_a\hat{n}^a = -\frac{1+\bar{n}^i h_{ij}\bar{n}^j}{S}.
\end{align}
Using Einstein's equation
\begin{align}
    T_{\mu\nu}=R_{\mu\nu}-\frac{R}{2}g_{\mu\nu}+\Lambda g_{\mu\nu},
\end{align}
the WEC simplifies to
\begin{align}
0&\leq T_{\mu\nu}n^\mu n^\nu 
\\
&= R_{\mu\nu}n^\mu n^\nu+\frac{R}{2}-\Lambda
\\
&= X \hat{n}_a\hat{n}^a + \bar{n}^i r_{ij}\bar{n}^j+\frac{R}{2}-\Lambda
\\
&=\bar{n}^i t_{ij} \bar{n}^j-\frac{X}{S}+\frac{R}{2}-\Lambda
\ \ \forall \bar{n}^j.
\end{align}
If we assume that the NEC is already satisfied everywhere, this just gives us 
\begin{align}
    0&\leq -\frac{X}{S}+\frac{R}{2}-\Lambda
    \\
    &=\frac{\bar{R}}{2}-\nablaII q -q^2 -\Lambda
    \label{WECforModel}
\end{align}
for $d=4$, using the general expression in \eqref{R4d} with $\beta=1$. 
The first two terms in \eqref{WECforModel} are total derivatives, so they will be non-positive \textit{somewhere}. The third term is negative \textit{everywhere}. So if it was only these terms, we could guarantee that the WEC will be violated. But the last term is positive for a negative cosmological constant, so by choosing $-\Lambda$ large enough we can always guarantee the WEC to be satisfied as long as the NEC is also satisfied. 
Essentially, a negative cosmological constant automatically violates the WEC (if we move it from the left to the right side of Einstein's equation and view it as a part of $T_{\mu\nu}$), so we can absorb any WEC violation from $T_{\mu\nu}$ by adjusting the value of $\Lambda$.  Clearly, a negative cosmological constant is not seen as problematic in the bottom-up holography community, despite the issue with the WEC. In our model building, as long as we remain agnostic about the exact value of $\Lambda$ and do not fix it, the WEC does thus not provide any additional benefit beyond the NEC.

\section{Scalar field as source of the spacetime curvature}
\label{sec::scalar}

In this section we assume that the spacetime is sourced by a scalar field, i.e.~that Einstein's equations for the bulk read
\begin{align}
R_{\mu\nu}-\frac{R}{2}g_{\mu\nu}=\nabla_\mu\varphi\nabla_\nu\varphi-g_{\mu\nu}\left(\frac{1}{2}(\nabla\varphi)^2+V(\varphi)\right).
\label{Einstein-Scalar-Eq}
\end{align}
Here, we have set the overall factor containing Newton's constant to one, and we have absorbed the cosmological constant into an overall shift of the potential $V(\varphi)$. Additionally, we will make the assumption that the scalar $\varphi$ should obey the same symmetries as the metric, i.e.~if $\mV$ is a Killing vector, then 
\begin{align}
    \mL_\mV\varphi = \mV^\mu\nabla_\mu \varphi =0. 
\end{align}
This implies that $\varphi$ can only be a function of the base--space coordinates, and 
\begin{align}
    \bar{\mL}_J\varphi = J^i\nablaII_i \varphi =0. 
    \label{JonVarphi}
\end{align}
Writing \eqref{Einstein-Scalar-Eq} in the warped product form analogous to \eqref{RmunuForm}, we find
\begin{align}
&\left(
\begin{array}{cc}
 X \eta_{ab}-\frac{R}{2}S\eta_{ab} & 0 \\
 0 & r_{ij} -\frac{R}{2}h_{ij} \\
\end{array}
\right)
\\
&=
\left(
\begin{array}{cc}
-S\eta_{ab}\left(\frac{1}{2}(\nabla\varphi)^2+V(\varphi)\right) & 0 \\
 0 & \nablaII_i\varphi\nablaII_j\varphi -h_{ij}\left(\frac{1}{2}(\nabla\varphi)^2+V(\varphi)\right) \\
\end{array}
\right)
\nonumber
\end{align}
From this, it is straightforward to show
\begin{align}
t_{ij}&\equiv r_{ij}-\frac{X}{S}h_{ij}=\nablaII_i\varphi\nablaII_j\varphi. 
\end{align}
This immediately implies $\det[t_{ij}]=0$ and because of \eqref{JonVarphi}, also 
\begin{align}
t_{ij}J^j=0\Rightarrow q^it_{ij}J^j=\phi=0. 
\end{align}
Consequently, scale invariance is automatically extended to full conformal invariance in this case (see section \ref{sec::phi0impliesconformal}). 

\ADS{While AdS$_3$ is a vacuum solution (with cosmological constant), this is not the case for \ADSxS which can indeed be sourced by a axion like scalar field, see e.g.~\cite{Balasubramanian:2013ux,Flory:2017mal} where AdS$_5\times$S${}^{1}$ arises as a special case of the models discussed there. Our results here thus clarify the connection between vanishing Weyl tensor, sourcing by a scalar field, and full conformal invariance which had been noticed in \cite{Flory:2017mal} but remained somewhat mysterious there. }

\section{One dimensional boundary}
\label{sec::1Dfiber}

The case of a one-dimensional boundary theory is very special, as there are no boosts and rotations present in the algebra. This will have significant consequences as we are about to see. Firstly, \eqref{algebras} simplifies to 
\begin{align}
\tcbhighmath[colback=green!0!white,colframe=green!75!black,size=small]{
  \begin{aligned}
    \text{Conformal invariance:}\hspace{5.95cm}
  \\
\tcbhighmath[colback=teal!0!white,colframe=teal!100!white,size=small]{
  \begin{aligned}
  \text{Scale without conformal invariance (SwCI):}\hspace{2.05cm}
  \\
  \begin{aligned}
\relax [D, P] &= P   \hspace{6.1cm}
  \end{aligned}
    \end{aligned}
}
\\
\relax [P, K] = -2 D\hspace{6.15cm}
\\
\relax [D, K] = -K \hspace{6.3cm}
    \end{aligned}
    }
    \label{algebras1d}
\end{align}
for the three generators $P_t\equiv P$, $D$, and $K_t\equiv K$. Now, starting with an assumed bulk Killing vector (and bulk coordinates $t,w,\theta$)
\begin{align}
P^\mu=\left(\begin{array}{c}
1\\
0 \\
0
\end{array}\right),
\label{P-type1d}
\end{align}
the algebra only implies 
\begin{align}
D^\mu=\left(\begin{array}{c}
-t+\mD(w,\theta)\\
J^w(w, \theta) \\
J^\theta(w, \theta)
\end{array}\right)
\label{D-type1d}
\end{align}
for scale invariance and
\begin{align}
K^\mu=\left(\begin{array}{c}
\mathcal{K}(w, \theta)+t^2-2t\mD(w,\theta)\\
-2 t J^w(w, \theta) \\
-2 t J^\theta(w, \theta)
\end{array}\right)
\label{Ktypes1d}
\end{align}
for full conformal invariance. Due to the absence of boosts in the algebra, these vectors are clearly less constrained than their higher dimensional counterparts \eqref{D-type} and \eqref{Ktypes}, as is evident from the appearance of the general function $\mD(w,\theta)$. Additionally, the condition $[D,K]=-K$ implies
\begin{align}
   0&= \mD\cdot J^i,
   \label{mKequationLong1dA}  
    \\
0&= -2\mD^2+2\mathcal{K} + J^\theta\partial_{\theta}\mathcal{K}+ J^w\partial_{w}\mathcal{K}
.
\label{mKequationLong1dB}  
\end{align}
These conditions would need to be satisfied if we wanted scale invariance to be extended to full conformal invariance. Suppose this was the case for a moment. Equation \eqref{mKequationLong1dA} would require either $J^i=0$ or $\mD=0$. 
Picking $J^i=0$ does not help, because then $D=(\mD(w,\theta)-t)\cdot P$. But if $\mV_1$ is a Killing vector and $\mV_2=\psi\mV_1$ with some scalar $\psi$, then $\mV_2$ can only be also a Killing vector if $\psi=const.$ which is clearly not possible here. So conformal invariance naively seems to require $\mD=0$, and then \eqref{mKequationLong1dB} simplifies to \eqref{mKequationLong}. 
Conversely, we see that $\mD\neq const.$\footnote{We require this instead of merely $\mD\neq 0$, because if $\mD=const.$, we can construct a new vector $D'=D-\mD\cdot P$ which is also a Killing vector if $P$ and $D$ are. } guarantees that scale invariance can not be extended to full conformal invariance. In that sense, the function $\mD$ which only appears for a one-dimensional boundary takes on the role that $\phi$ played in higher dimensions.
\\

To find a three-dimensional bulk model with the isometry group spanned by $P$ and $D$, we would make the generic ansatz 
\begin{align}
ds^2=-S(w,\theta)dt^2+h_{ij}(w,\theta)dy^idy^j+k_{i}(w,\theta)dt dy^i
\label{Metric_ansatzD1}
\end{align}
with $y^i\in\{w,\theta\}$. The fact that all metric components are independent of the coordinate $t$ guarantees that \eqref{P-type1d} is a Killing vector. This ansatz is more general than the warped product form \eqref{Metric_ansatz} due to the presence of the terms $k_{i}(w,\theta)dtdy^i$\footnote{Hence, technically we shouldn't use the terms "fiber" and "base space" anymore in this context. } -- these would be inconsistent with Killing vectors corresponding to boosts or rotations, which however do not exist in \eqref{algebras1d}. As both \eqref{Metric_ansatzD1} and \eqref{D-type1d} differ qualitatively from what we have worked with in the main text, our methods developed there do not apply straightforwardly to the case of one boundary dimension, and so the validity of our main conjecture in box \ref{B1} for this case is left for future research.

\section{Higher dimensional fiber}
\label{sec::higherDfiber}

Throughout the paper, our strategy has been to take problems in the full bulk and reformulate them as problems in the two-dimensional base space. To show that our results hold for any fiber dimension $n\geq2$ (while keeping the base space two-dimensional), we thus have to check to what degree the base space equations and arguments that we arrived at depend on the fiber dimension. This is exactly what we will do in this section. 
We start from equation \eqref{Metric_ansatz} with coordinates $x^a\in\{t,x^1,\dots,x^{n-1}$) and $y^i\in\{w,\theta\}$. Due to the constraints imposed by the algebra \eqref{algebras}, the potential generator of scale transformations takes the form (equation \eqref{D-type} in $n=2$)
\begin{align}
D^\mu=\left(\begin{array}{c}
-t\\
-x^1\\
\vdots \\
-x^{n-1} \\
J^w(w, \theta) \\
J^\theta(w, \theta)
\end{array}\right).
\label{D-typeD}
\end{align}
and the generators of special conformal transformations would have to read (equation \eqref{Ktypes} in $n=2$)
\begin{align}
(K_0)^\mu=\left(\begin{array}{c}
\mathcal{K}(w, \theta)+t^2+(x^1)^2+\dots\\
2 t x^1 \\
\vdots\\
2 t x^{n-1} \\
-2 t J^w(w, \theta) \\
-2 t J^\theta(w, \theta)
\end{array}\right),\ \ 
(K_1)^\mu=\left(\begin{array}{c}
-2 t x^1 \\
\mathcal{K}(w, \theta)-t^2-(x^1)^2-\dots\\
\vdots\\
-2 x^1 x^{n-1} \\
2 x^1 J^w(w, \theta) \\
2 x^1 J^\theta(w, \theta)
\end{array}\right),\dots\ .
\label{KtypesD}
\end{align}
Equations \eqref{mKequationLong}, \eqref{Sequation}, \eqref{Fequation} and \eqref{Kequation} all remain valid for general $n\geq2$. 
\\

Let us now investigate the Weyl tensor. In $d=n+2$ bulk dimensions, the Weyl tensor is defined as \cite{Stephani:2003tm}
\begin{align}
C_{\alpha\beta\gamma\delta}\equiv R_{\alpha\beta\gamma\delta}
-\frac{1}{d-2}\left(g_{\alpha[\gamma }R_{\delta ]\beta}-g_{\beta[\gamma }R_{\delta ]\alpha}\right)
+\frac{1}{(d-1)(d-2)}Rg_{\alpha[\gamma }g_{\delta ]\beta}
\end{align}
generalising \eqref{WeylAndReducedWeyl}. We still find the basic structure of \eqref{WeylForm},
\begin{align}
C_{\alpha\beta\gamma\delta}=\phi\, \tilde{C}_{\alpha\beta\gamma\delta},
\end{align}
with the same definition of $\phi$ \eqref{PhiEq1} \cite{Nozawa:2008rjk}, 
but now $C_{\mu \nu \alpha \beta} C^{\mu \nu \alpha \beta} = f(d) \phi^2$ with some function $f(d)$ as the dimension dependent numerical factor. As there is no significant dependence on the dimension that has appeared in our equations so far, we can conclude that the derivations of sections \ref{sec::ConformalConditions} - \ref{sec::phi0impliesconformal} and thus the result in box \ref{B0} remain valid for any $n$.

Next we proceed to investigate the NEC, and the arguments leading to the proof given in section \ref{sec::full_proof}. Due to the warped product form of the metric, \eqref{RmunuForm} remains valid, but now we find \cite{Nozawa:2008rjk}
\begin{align}
    X^{(d)}&= -[\sqrt{S} \boxII\sqrt{S}+ (d-3)(\nablaII \sqrt{S})^2],
        \label{XdefD}
        \\
r_{ij}^{(d)} &= \frac{\bar{R}}{2} h_{ij} - \frac{(d-2)}{\sqrt{S}}\nablaII_i\nablaII_j\sqrt{S}.
\end{align}
 Consequently, the NEC is related to the semi-positiveness of the tensor 
\begin{align}
t_{ij}^{(d)}&\equiv r_{ij}^{(d)}-\frac{X^{(d)}}{S}h_{ij}
\\
&=\left[\frac{\bar{R}}{2}+\frac{\boxII\sqrt{S}}{\sqrt{S}}+(d-3)\frac{(\nablaII \sqrt{S})^2}{S} \right]h_{ij}- \frac{(d-2)}{\sqrt{S}}\nablaII_i\nablaII_j\sqrt{S}
\\
&=\left[\frac{\bar{R}}{2}+\nablaII q + (d-2)q^2 \right]h_{ij}- (d-2)\left[\nablaII_i q_j + q_i q_j\right]
\label{tijqD}
\end{align}
with 
\begin{align}
\text{tr}[t_{ij}^{(d)}]&=\bar{R}+(d-2) q^2-(d-4)\nablaII q ,
\\
\det[t_{ij}^{(d)}]&=
\left[\frac{\bar{R}}{2}+\nablaII q+(d-2)q^2\right]^2
+\left[\frac{\bar{R}}{2}+\nablaII q+(d-2)q^2\right](2-d)(\nablaII q+q^2) \notag
\\
&+\frac{(d-2)^2}{2}\left[(\nablaII q+q^2)^2-(\nablaII_i q_j + q_i q_j)(\nablaII^i q^j + q^i q^j)\right].
\label{DETtijqD}
\end{align}

As this tensor depends on the dimension, we have to carefully check whether our main conjecture of box \ref{B1} still holds. We will not go through all the special case proofs of section \ref{sec::special_cases}, however we note that the crucial result \eqref{qtJ} still holds:
\begin{align}
    q^i t_{ij}^{(d)} J^j = \frac{\phi}{2},
\end{align}
independently of the dimension $d$. Thus, the argument given for the special case investigated in section \ref{sec::q||J} still holds trivially. 
\\

To proceed with a general proof in this case, we follow the same strategy as in section \ref{sec::full_proof}, keeping $\beta=1$ without loss of generality. First of all, we can verify that $\det[t_{ij}^{(d)}] = 0$ when $\phi = 0$, using the specific vector $q^i$ given in \eqref{q_for_phi0}. Moreover, we observe that the vector $J^i$ played a crucial role in proving our theorem in section \ref{sec::full_proof}. Therefore, using \eqref{tijqD}, we can write: 
\begin{align}
    J^i t_{ij}^{(d)} J^j &= \left(\frac{\bar{R}}{2} + \nablaII q \right) J^2 - (d-2) \left[(1 - q^2 J^2) + J^i J^j \nablaII_i q_j\right] \notag \\
    &= \frac{\phi}{2} J^2 - (d-2) \left[(1 - q^2 J^2) + \frac{1}{2} q^i \nablaII_i J^2\right]\,.
\end{align}
Furthermore, we can use the form of $q^i$ from \eqref{q_for_phi0} for the $\phi = 0$ case to evaluate:
\begin{align}
    J^i t_{ij}^{(d)} J^j = - (d-2) \left[\left\{1 - \left(\frac{1}{J^2} + \frac{1}{4(J^2)^2} \nablaII_i J^2 \nablaII^i J^2\right) J^2 \right\} + \frac{1}{4J^2} \nablaII^i J^2 \nablaII_i J^2 \right] = 0\,.
\end{align}
We can consider a perturbation around $\phi = 0$, as in \eqref{q_ansatz_gamma}, and examine the signatures of NEC violation using the vector $J^i$. In this case, for $\phi \neq 0$, we obtain:
\begin{align}
\label{JtdJ}
    J^i t_{ij}^{(d)} J^j = \nablaII_i (\gamma^i J^2) + 2 \gamma^2 J^2 + \frac{(d-4)}{2} (2 \gamma^2 J^2 + \gamma^i \nablaII_i J^2)\,.
\end{align}
As a consistency check, we can take $d = 4$ and verify that we recover \eqref{JtJsimplified}. Again, we can adopt a suitable coordinate system as detailed in section \ref{sec::full_proof} and arrive at:
\begin{align}
    J^i t_{ij}^{(d)} J^j &= \frac{J^4}{h} \partial_\theta W + \frac{W}{\sqrt{h}} \partial_\theta \frac{J^4}{\sqrt{h}} + \frac{2 J^4}{h} W^2 + \frac{(d-4)}{2} \left( \frac{W}{h} J^2 \partial_\theta J^2 + \frac{2 J^4}{h} W^2 \right)
\end{align}
We notice that the additional terms appearing in the above expression due to the higher-dimensional fiber are all polynomial in $W(\theta)$. While proving our conjecture in box \ref{B1} for general cases in section \ref{sec::full_proof}, we observed that the leading-order contribution to $J^i t_{ij} J^j$ around points where $W(\theta)=0$ arises from the derivative term $\partial_\theta W$ for both analytic and non-analytic functions $W(\theta)$, which leads to NEC violation. We clearly see that the higher-dimensional contributions to $J^i t_{ij}^{(d)} J^j$ are subleading, and our proof in section \ref{sec::full_proof} extends to arbitrary dimensions ($n \geq 2$) of the fiber.

\bibliographystyle{JHEP}
\bibliography{References}

\end{document}